\documentclass{ieeeaccess}
\usepackage{cite}
\usepackage{amsmath,amssymb,amsfonts}
\usepackage{algorithmic}
\usepackage{graphicx}
\usepackage{textcomp}
\def\BibTeX{{\rm B\kern-.05em{\sc i\kern-.025em b}\kern-.08em
    T\kern-.1667em\lower.7ex\hbox{E}\kern-.125emX}}
    
\usepackage[utf8]{inputenc}
\usepackage[T1]{fontenc}

 \usepackage[hyphens]{url}
 \usepackage{hyperref}
 \hypersetup{breaklinks=true}
 
 \usepackage{eurosym}
 \usepackage{authblk}
 \usepackage{cite}
 \usepackage{epstopdf}
 \usepackage{tikz,lipsum}
 \NewSpotColorSpace{PANTONE}
   \AddSpotColor{PANTONE} {PANTONE3015C} {PANTONE\SpotSpace 3015\SpotSpace C} {1 0.3 0 0.2}
  \SetPageColorSpace{PANTONE}
 \usepackage{caption}
 \usepackage{float}
 \usepackage{mathrsfs}
 \usepackage{makecell}
 \usepackage{mathrsfs}
 \usepackage{color}
 \usepackage{multirow}
 \usepackage{lscape}
 \usepackage{xcolor,colortbl}
 \usepackage{array}
 \DeclareUnicodeCharacter{20AC}{\euro}
 \usepackage{pifont}% http://ctan.org/pkg/pifont
 \usepackage{forest}
 \forestset{
   dir tree/.style={
    for tree={
      parent anchor=south west,
      child anchor=west,
      anchor=mid west,
      inner ysep=1pt,
      grow'=0,
      align=left,
      edge path={
        \noexpand\path [draw, \forestoption{edge}] (!u.parent anchor) ++(1em,0) |- (.child anchor)\forestoption{edge label};
      },
      if n children=0{}{
        delay={
          prepend={[,phantom, calign with current]}
        }
      },
      fit=band,
      before computing xy={
        l=2em
      }
    },
  }
}
\usepackage{adjustbox}

\definecolor{LightCyan}{RGB}{0.88,1,1}
\definecolor{antiquewhite}{RGB}{0.98, 0.92, 0.84}
\definecolor{cottoncandy}{RGB}{1.0, 0.74, 0.85}
\definecolor{wheat}{RGB}{0.96, 0.87, 0.7}

\newcommand{\specialcelll}[2][l]{%
\begin{tabular}[#1]{@{}l@{}}#2\end{tabular}
}
\def\BibTeX{{\rm B\kern-.05em{\sc i\kern-.025em b}\kern-.08em
    T\kern-.1667em\lower.7ex\hbox{E}\kern-.125emX}}

 \usepackage{tikz}
 \newcommand*\emptycirc[1][1ex]{\tikz\draw (0,0) circle (#1);} 
\newcommand*\halfcirc[1][1ex]{%
  \begin{tikzpicture}
  \draw[fill] (0,0)-- (90:#1) arc (90:270:#1) -- cycle ;
  \draw (0,0) circle (#1);
  \end{tikzpicture}}
\newcommand*\fullcirc[1][1ex]{\tikz\fill (0,0) circle (#1);} 
 
\begin{document}
 \history{Date of publication xxxx 00, 0000, date of current version xxxx 00, 0000.}
 \doi{10.1109/ACCESS.2017.DOI}

\title{Reliable and Resilient AI and IoT-based Personalised Healthcare Services: A Survey}
\author{\uppercase{Najma Taimoor}\authorrefmark{1},
        \uppercase{Semeen Rehman}\authorrefmark{2},
}

\address[1,2]{Institute of Computer Technology, Vienna University of Technology, Austria}

\markboth
{Najma \headeretal: Reliable and Resilient AI and IoT-based Personalised Healthcare Services: A Survey}
{Najma \headeretal: Reliable and Resilient AI and IoT-based Personalised Healthcare Services: A Survey}

\corresp{Corresponding author: Najma Taimoor (e-mail: e1155891@student.tuwien.ac.at).}
\begin{abstract}
Recent technological (e.g., IoT, 5G), and economic (e.g., UN 2030 Sustainable Development Goals) developments have transformed the healthcare sector towards more personalized and IoT-based healthcare services. These services are realized through control and monitoring applications that are typically developed using artificial intelligence (AI)/machine learning (ML) based algorithms, that play a significant role to highlight the efficiency of traditional healthcare systems. Current personalized healthcare services are dedicated in a specific environment to support technological personalization  (e.g., personalized gadgets/devices). However, they are unable to consider different inter-related health conditions, leading to inappropriate diagnosis and affect sustainability and the long-term health/life of patients. Towards this problem, the state-of-the-art Healthcare 5.0 technology has evolved that supersede previous healthcare technologies. The goal of healthcare 5.0 is to achieve a fully autonomous healthcare service, that takes into account the interdependent effect of different health conditions of a patient. This paper conducts a comprehensive survey on personalized healthcare services. In particular, we first present an overview of key requirements of comprehensive personalized healthcare services (CPHS) in modern healthcare Internet of Things (HIoT), including the definition of personalization and an example use case scenario as a representative for modern HIoT. Second, we explored a fundamental three-layer architecture for IoT-based healthcare systems using both AI and non-AI-based approaches, considering key requirements for CPHS followed by their strengths and weaknesses in the frame of personalized healthcare services. Third, we highlighted different security threats against each layer of IoT architecture along with the possible AI and non-AI-based solutions. Finally, we propose a methodology to develop reliable, resilient, and personalized healthcare services that address the identified weaknesses of existing approaches.
\end{abstract}


\titlepgskip=-15pt

\maketitle

\section{Introduction}
\label{sec:intro}
\PARstart{C}{ontemporary} social and economic developments around the globe e.g., UN Sustainable Agenda 2030~\cite{un2030}, are aiming to extend life expectancy for all humans by improving their physical and mental health, and well-being. Current technological advancements have helped to achieve this agenda in real-time. To this end, various dedicated technical initiatives have been introduced, for instance, one such initiative is "Healthcare 5.0" which has been developed as a result of the emergence of digital wellness and a new digital standard for healthcare services~\cite{Marek2017}. Furthermore, recent development in technology has enabled remote and automatic monitoring of healthcare services through medical devices that aims to monitor various health conditions of a patient. Importantly, these devices are dedicated to specific health conditions and work independently. For instance, blood pressure monitor aims to monitor only heart-related health conditions, insulin pump only regulates diabetic health conditions through maintaining the right level of blood insulin, to name a few.

\begin{figure}[ht]
    \centering
    \includegraphics[scale=0.60]{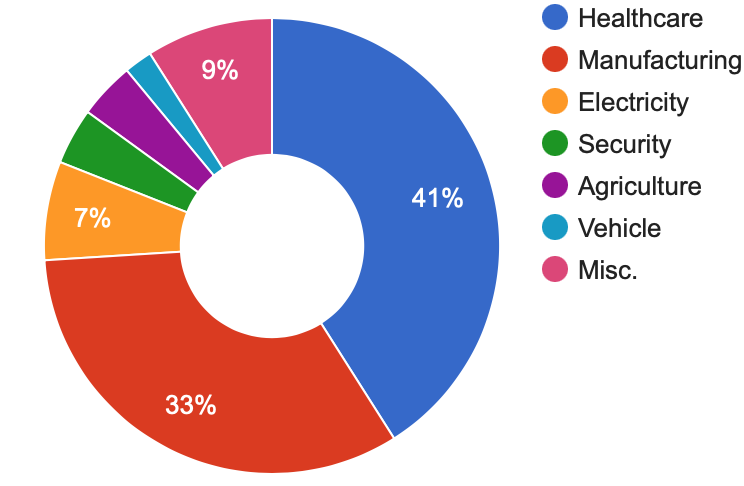}
    \caption{Economic Impact of IoT Devices and Applications}
    \label{fig:ecoimpact}
\end{figure}

%\subsection{History of Healthcare}
Prior to Healthcare 5.0, "Healthcare 4.0" had emerged from Industry 4.0, that had transformed the healthcare sector into more digital during the past decade. For instance,  x‐rays and magnetic resonance imaging (MRI) have transformed into computer tomography (CT) and ultrasound scans to electric medical records~\cite{Bercovich2018}. These devices are user-centric that are used by caretakers/ medical practitioners to monitor and treat medical conditions of patients, in preventive care, and for well-being solutions. With the recent advent of technologies (e.g., Industrial Internet of Things IIoT, Industrial cyber-physical systems ICPS), the use of IoT devices and applications is exponentially growing. According to \cite{giri2017} by the end of 2021, the IoT devices and applications are expected to reach 212 billion, whose major use is in healthcare (\textasciitilde 41\%) as shown in Figure~\ref{fig:ecoimpact}. Furthermore, as reported by Grand View Research~\cite{gvr2019}, the healthcare IoT market is expected to be worth USD 534.3 billion by 2025.
% The IoT market in healthcare is predicted to grow up to 188 billion US dollars by 2024, at a Compound Annual Growth Rate (CAGR) of 27.6\% during the forecast period~\cite{IoTmarket2020}.
%%%%%%%%%%%%%%%%%%%
\begin{table*}[ht]
\caption{Perspective-based Healthcare System/Infrastructure Components}
\centering
%\begin{tabular}{|l | l |l |}
\begin{tabular}{|>{\centering\arraybackslash}m{3cm} |>{\centering\arraybackslash}m{6.2cm} |>{\centering\arraybackslash}m{6.2cm}|}
\hline
\hline
\rowcolor{gray!20}
\diaghead{\theadfont Diag Columnmn Head II}%
  {\bf{Perspective}}{\bf{Components}} & \textbf{Hardware} & \textbf{Software}\\
 \hline \hline
\textbf{Technological} &  \specialcelll{Personalized medical devices and gadgets \\(e.g., Smart shirts and watches)} & \specialcelll{Technology-driven applications \\(e.g., AI/ML-based patient specific control \\and monitoring applications)} \\
\hline

\textbf{Clinical} & \specialcelll{Underlying sensing infrastructure\\ (e.g., Bio and pressure sensors)} & \specialcelll{Clinical process-driven applications \\(e.g., AI/ML-based diagnosis and treatment \\ applications)}\\
\hline
\end{tabular}
\label{table:ctperspective}
\end{table*}
  %%%%%%%%%%%%%%%%%%%%%%%%%%%%%%%%%%%

%\subsection{Healthcare 5.0 specifics}
The principle goal of technology-driven applications in healthcare is to support healthcare operations by automatically controlling various health conditions of patients through continuous and remote monitoring of the conditions~\cite{ramson2020,zhan2020,alshorman2020}. Following the principle, Healthcare 5.0 aims to support various objectives, i.e., (i) reliable, (ii) resilient, and (iii) personalized healthcare services in real-time. In the following, we discuss our taxonomy that characterizes the aforementioned healthcare objectives.
\begin{itemize}
%\subsubsection{Reliability}
    \item \textbf{Reliability:} Analogous to the definition of reliability by IEEE Standards~\cite{g1990}, we consider the reliability of healthcare services as the ability of the services to perform their required functions consistent with stated conditions (aka specification) for a specified time as shown in Figure~\ref{fig:reliability}.
    Healthcare 5.0 supports automatic monitoring and control of underlying health conditions of patients. Therefore, these services must be extremely reliable (i.e., they must perform required operations as expected under normal and stated conditions). Otherwise, the services may fail to operate as expected leading to severe consequences, including threatening the lives of patients. To establish reliability, various efforts~\cite{vyas2019,vergutz2020,levashenko2016} have identified different factors (e.g., battery, memory, computational power, enhancement of QoS) that can improve the reliability of the healthcare services by using modern solutions e.g., blockchain-based technologies.
%%%%%%%%%%%%%%%%%%%%%%%%
\begin{figure}[ht]
    \centering 
    \includegraphics[scale=0.75]{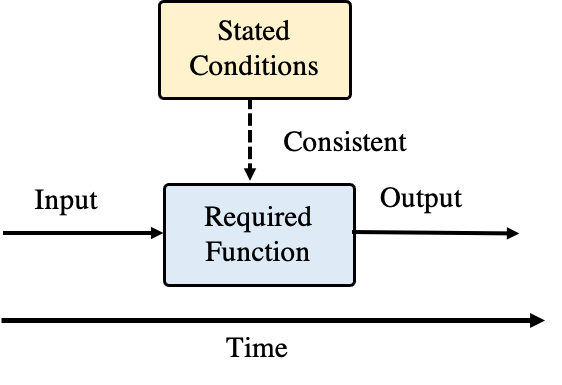}
    \caption{Reliability}
    \label{fig:reliability}
\end{figure}
% %%%%%%%%%%%%%%%%%%%%%%%%
%\subsubsection{Resilience}
    \item \textbf{Resilience:} We adapt the definition of resilience from different interpretations~\cite{firesmith_2019, Avizienis2004}, as the ability of healthcare service to continue its required function in the face of adverse operating conditions. To this end, a resilient service first detects an adverse condition (e.g., fault, error, bug, and cyber threat) that malfunctions it, then recovers the required function of the service by mitigating the impact of the adverse condition as shown in Figure~\ref{fig:resilience}. 
    Healthcare 5.0 services need to be resilient to support continuous monitoring of health conditions, i.e., to continuously deliver correct functionality and availability of the services to the user despite adverse environmental conditions, internal faults of the system such as hardware and software defects, cybersecurity threats, and vulnerabilities, excessive loads, age, and wear, and degraded communications~\cite{firesmith_2019}. Any disruption of the services due to intentional or accidental incidents may also threaten patients' lives. To establish resilience, various efforts~\cite{zhang2020,rajmaki2018}
have recognized potentially vulnerable functions (e.g., failure of software, hardware, cloud services, and communication devices) that may hinder the resilience of healthcare attributes and proposed different ways to recover from them (e.g., post-event automatic recovery of the vulnerabilities like Moving Target Defense (MTD)~\cite{sood2018}, and blockchain-based solutions~\cite{safavi2018}).

    \item \textbf{Personalization:} Personalized healthcare services typically operate in the most strict mode by supporting customization of a specific health condition under specific conditions~\cite{zhang2019,Chauveau2018}. This does not work in practice because patients with long-term health conditions often have multiple health conditions. Therefore, we realize personalization of healthcare service as the ability of the service that provides determinant-based (e.g., genetics, behavior, environmental and physical influences, medical care, and social factors) optimization of multiple health conditions of a patient as shown in figure~\ref{fig:personalization}. The optimization extends the life expectancy of the patient by reducing the side effects of various health conditions.
    %%%%%%%%%%%%%%%%%%%%%%%%%%%%%%%%%
\begin{figure}[ht]
    \centering 
    \includegraphics[scale=0.60]{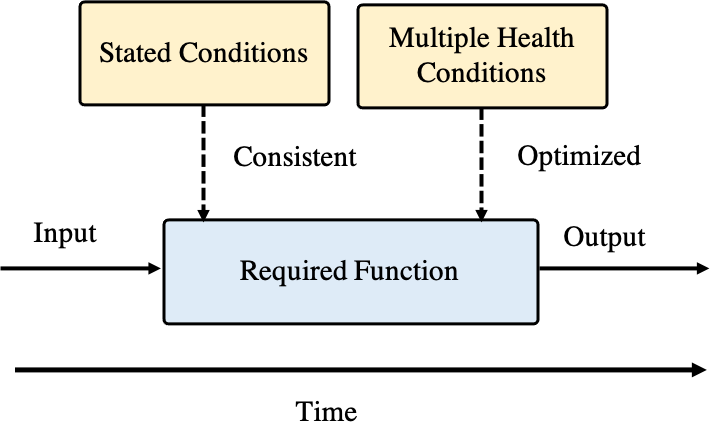}
    \caption{Personalization}
    \label{fig:personalization}
\end{figure}
%%%%%%%%%%%%%%%%%%%%%%%%%%%%%%%%%
 %%%%%%%%%%%%%%%%%%%%%%%%%
\begin{figure*}[ht]
    \centering 
    \includegraphics[scale=0.65]{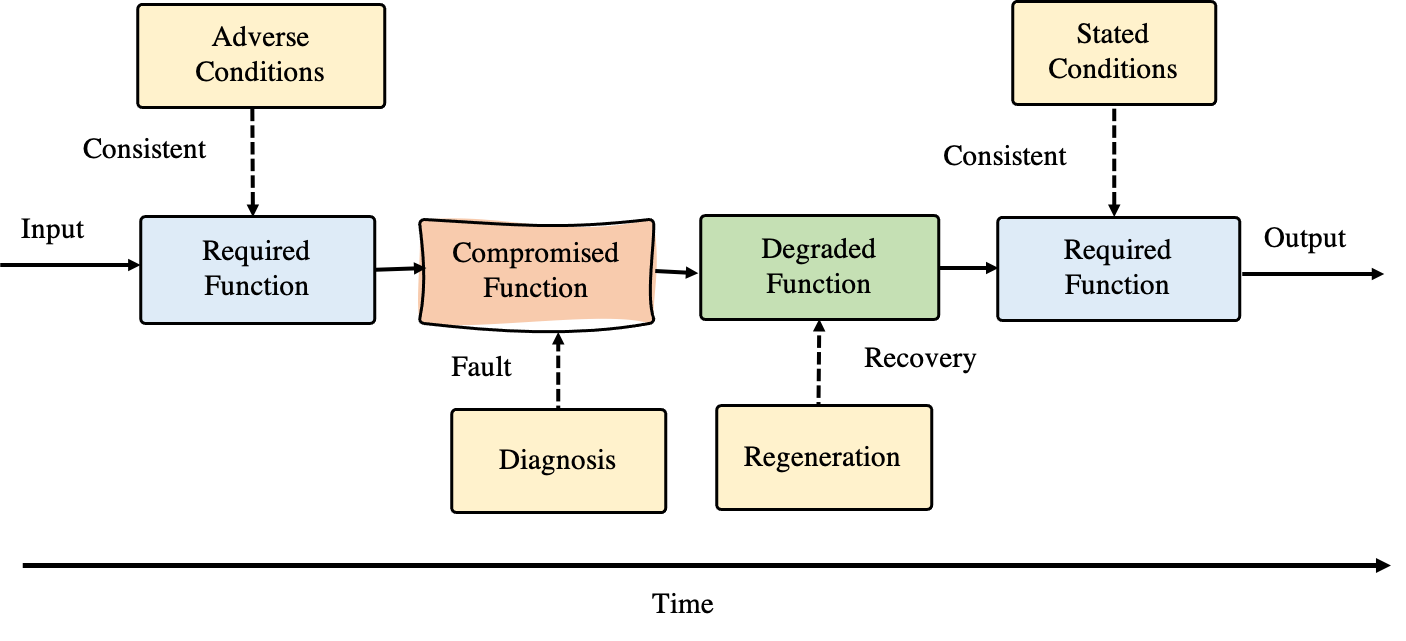}
    \caption{Resilience}
    \label{fig:resilience}
\end{figure*}
%%%%%%%%%%%%%%%%%%%%%%%%%
    
%Another vital goal of technology-driven healthcare services is to extend the life expectancy of patients through resolving their various health conditions. 
Based on our taxonomy, we extend Healthcare 5.0 to support personalized healthcare services that aim to optimize clinically inter-related multiple health conditions of a patient. As sketched in Table~\ref{table:ctperspective}, the current infrastructure (i.e., software and hardware) of healthcare services supports complementary technological and clinical personalization. Former approaches aim to develop personalized medical gadgets (e.g., customized hearing aid, artificial placenta~\cite{placenta2021}, wearable insoles~\cite{insoles2019}) and applications (e.g., Basal and bolus insulin settings in insulin-pump~\cite{frida2021}, fitbit+~\cite{almogbil2020}) for handling specific technical needs of a particular health condition. Latter approaches aim to develop personalized medical technologies (e.g., enzyme-based biosensors for sweat analysis~\cite{elena2020}, antibody-based sensor for rapid detection of avian coronavirus~\cite{weng2018}) and applications (e.g., ML-based algorithms~\cite{he2019}, clinical-parameter
-based configuration of insulin-pump~\cite{berget2020}) for handling specific biological effects of a particular health condition.

\end{itemize} 
   
As discussed above, current approaches to establish reliability, resilience, and personalization of healthcare services fail to support continuous monitoring of health conditions in Healthcare 5.0 mainly because they are either health condition-specific, or environment-specific, or agent-specific. Moreover, the emergence of new IoT medical devices manufactured by different vendors and their communication through different protocols have made the task of establishing the above-mentioned requirements more challenging. Therefore, in this survey, we systematically study and analyze various approaches that establish key requirements of the services and identify gaps that help to establish Healthcare 5.0 services in practice.

 \textbf{Our Contributions:}
% In contrast to the existing surveys, this paper targets to present an in-depth and architecture survey and analysis of applications, Things (e.g., IIoT/ICPS), and communication for today's personalized healthcare systems. 
To the best of our knowledge, this is the first attempt that identifies and studies key requirements of HIoT to support clinical personalization of healthcare services (we call it CPHS) in contrast to the existing ones that support only technological personalization. We summarize our contributions as follows:
\begin{enumerate}
    \item We defined novel comprehensive personalized healthcare services (CPHS) by contextualizing HIoT (Healthcare 5.0) to support clinical personalization
    \item We identified key functional (reliability, resilience, and personalization) and non-functional (constraints on reliability, resilience, and personalization) requirements of the CPHS
    \item We introduced an example use case scenario as a representative for modern HIoT
    \item We explored and analyzed (strengths vs weaknesses) various AI and non-AI-based approaches to establish the identified requirements of HIoT (e.g., reliability, resilience, and personalization) in each layer of the reference architecture (i.e., things, communication, and application layer) under normal and hostile conditions (e.g., security threats).

    % \item We explored and analyzed various security threats and their potential mitigation at different layers of the reference architecture.
    % % \item We identify the future research directions in the personalized healthcare industry that exploits the advantages of the novel applications and technologies presented in the discussion.
    \item Finally, we discussed our proposed methodology to provide state-of-the-art reliable, resilient, and real-time comprehensive personalized healthcare services that address the identified weaknesses of the existing approaches.
\end{enumerate}
The overall logical organization of the paper is sketched in Figure~\ref{fig:paperstructure}. The rest of the paper is organized as follows. Section~\ref{sec:sm} explains our survey methodology. Section~\ref{sec:thc} introduces the transformation of healthcare technology, personalized healthcare, and their key operational requirements along with the example use case scenario. Section~\ref{sec:r_architecture} gives an overview of a reference architecture for IoT-based healthcare systems, their different layers, and recent efforts to establish requirements of the modern personalized healthcare services. Section~\ref{sec:s_threats} provides an overview of security threats at IoT layers and types of attacks on each layer, while Section~\ref{sec:psolution} introduces our proposed solution. Finally, Section~\ref{sec:conclusion} concludes our paper.
\section{Survey Methodology}\label{sec:sm}
We have performed a survey in a very systematic way by identifying top research findings in the domain of IoT and healthcare. We extracted information and summarised the current literature according to the guidelines suggested by Kitchenham~\cite{kitchenham2009systematic, keele2007guidelines}. Our approach follows a defined sequence of steps using a systematic literature review (SLR)~\cite{okoli2010guide}, started from downloading related research articles, reports, and thesis using keyword-based query-searching mentioned in Table~\ref{tab:search-criteria}. Our initial web search found more than 30000 papers from different data sources, e.g., Springer, IEEE, Science Direct, Academic medicine, ACM. Papers were accessed mainly from google scholar. Then our refined search based on related keywords (e.g., healthcare systems, IoT, healthcare 5.0, medicine 4.0, personalized healthcare) found approx. 200 papers. Our final refinement eliminated further 40 papers that were highly theoretical and from non-scientific sources. The remaining papers also included surveys in healthcare, however, most of these surveys~\cite {Dridi2017,Amin2019 , Alfian2018} are focused on monitoring specific health conditions (e.g., cardiovascular, breathing problems, diabetes, etc.) in a specific environment ~\cite{Abdulrauf2018,Ahmed2015} (e.g., smart home, assisted living environment, elderly homes,) for specific healthcare services (e.g., wellness services and health status)~\cite{ALI2016,ma2017} and for specific agents (e.g., caretakers, doctors) using modern infrastructure (e.g., IoT devices, 5G network, and electronic healthcare).

% Furthermore, contemporary approaches have addressed only one specific requirement of healthcare systems, e.g., reliability, resilience, or personalization. All the above approaches view personalization in a way that is not practical. Since these approaches do not provide a unified view of personalization, therefore, such a view can not be fully realized automatically and thus fail to achieve the automation goals\cite{Mohanta2019} of healthcare 5.0.

%%%%%%%%%%%%%%%%%%%%%%
\begin{figure*}[ht]
    \centering 
    \includegraphics[scale=0.55]{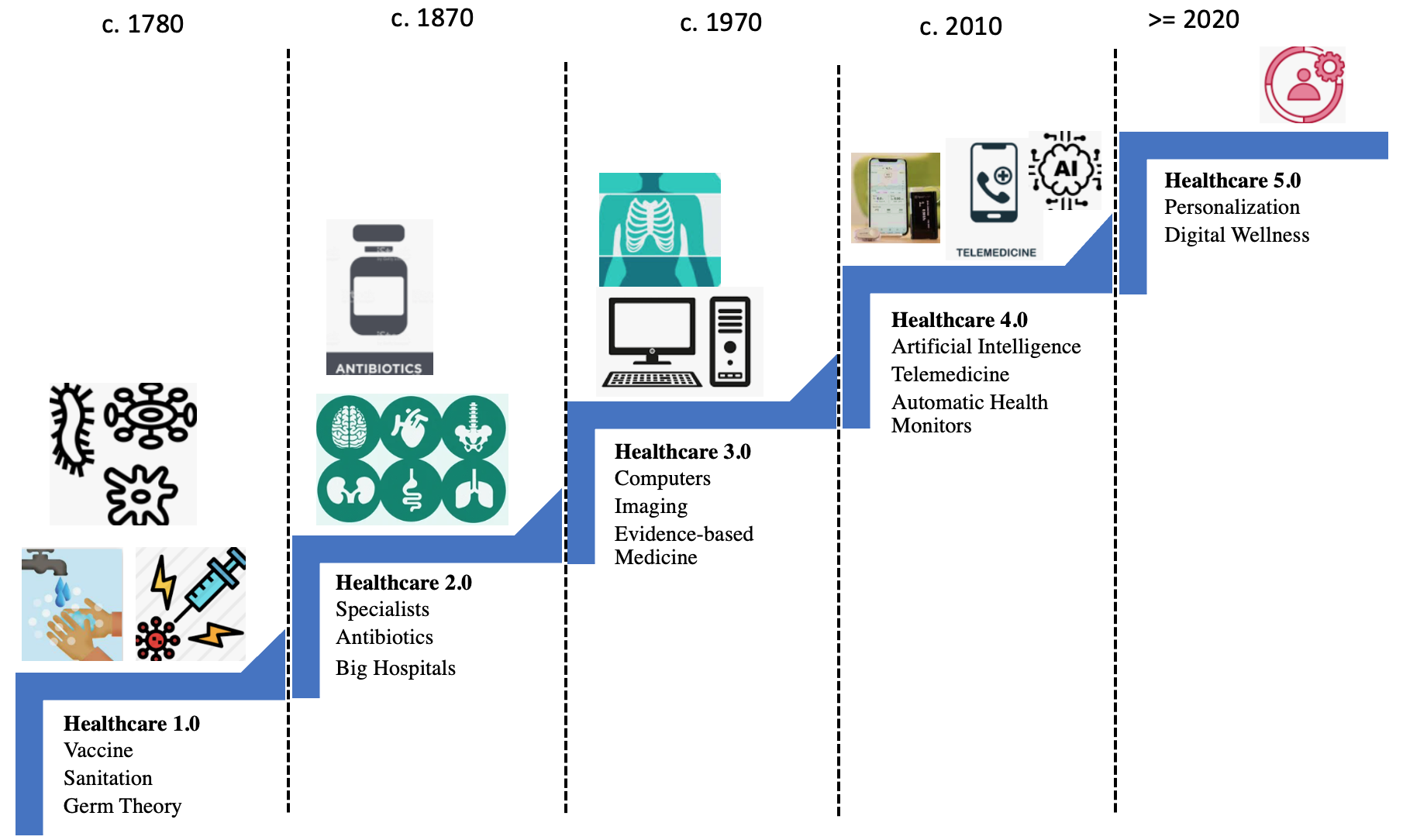}
    \caption{Timeline of Healthcare evolution}
    \label{fig:h1-5}
\end{figure*}
%%%%%%%%%%%%%%%%%%%%%%%%%%%%%

%%%%%%%%%%%%%%%%%
\begin{table}[h]
\caption{
Keywords}
\label{tab:search-criteria}
\centering
\begin{tabular}{|p{0.075\textwidth}|p{0.035\textwidth}|p{0.3\textwidth}|} 
\hline 
\multirow{9}{*}  &  & personalization\\
\cline{3-3}
 &  & IoT \textbf{OR} HIoT\\
\cline{3-3}
&  & Security \textbf{OR} Reliability \\
\cline{3-3}
& & Reliability \textbf{AND} Things \textbf{OR} sensors \\
\cline{3-3}
& & Reliability \textbf{AND} Things \textbf{OR} AI/ML \\
\cline{3-3}
& & Reliability \textbf{AND} communication \textbf{OR} network \\
\cline{3-3}
& & Reliability \textbf{AND} application \textbf{OR} software \\
\cline{3-3}
& & (Resilience \textbf{AND} Things) \textbf{AND} sensors \\
\cline{3-3}
& & (Resilience \textbf{AND} communication \textbf{OR} network \\
\cline{3-3}
& & (Resilience \textbf{AND} communication \textbf{OR} AI/ML \\
\cline{3-3}
& & (Resilience \textbf{AND} application \textbf{OR} software \\
\cline{3-3}
Healthcare & \textbf{AND} & (security \textbf{OR} privacy) \textbf{AND} threats\\
\cline{3-3}
& & Personalization \textbf{AND} Things \textbf{OR} sensors \\
\cline{3-3}
& & Personalization \textbf{AND} communication \textbf{OR} network \\
\cline{3-3}
& & Personalization \textbf{AND} application \textbf{OR} software \\
\cline{3-3}
& & "IoT layers" \textbf{OR} "manufacturing industry"  \\
\cline{3-3}
& & sensors \textbf{AND} reliability\\
\cline{3-3}
& & sensors \textbf{OR} sensor types \textbf{AND} healthcare\\
\cline{3-3}
& & Healthcare 5.0 \textbf{OR} Medicine 4.0 \textbf{AND} healthcare\\
\cline{3-3}
& & Healthcare 5.0 \textbf{OR} AI/ML \textbf{AND} Personalized healthcare\\
\cline{3-3}
& & wearables \textbf{AND} healthcare \\\hline
\end{tabular}
\end{table}
%%%%%%%%%%%%%%%%%

\begin{figure}
    \centering
    \begin{forest}
  dir tree
  [\textbf{Reliable Resilient and Personalized Healthcare 5.0}
    [\textsection \ref{sec:intro} Introduction]
    [\textsection \ref{sec:sm} Survey Methodology]
    [\textsection \ref{sec:thc} Transformation of Healthcare Technologies
      [\textsection \ref{subsec:ph} Personalized Healthcare (p-Health)
        % [\textsection \ref{subsec:cphs} Comprehensive p-Health Services (CPHS)]
        [Example Use Case Scenario]
      ]
      [\textsection \ref{subsec:keyreqs} Key Requirements of CPHS
      [Reliability]
      [Resilience]
      [Personalization]
      ]
    ]
    [\textsection \ref{sec:r_architecture} A Reference Architecture
    [\textsection \ref{subsec:tl} Things Layer
    [Overview/State-of-the-art]
    [Analysis of CPHS Requirements
    ]
    ]
    [\textsection \ref{subsec:cl} Communication Layer
    [Overview/State-of-the-art]
    [Analysis of CPHS Requirements
    ]
    ]
    [\textsection \ref{subsec:al} Application Layer
    [Overview/State-of-the-art
    [Support for p-Health]
    [Healthcare Modeling]
    [Healthcare Monitoring]
    ]
    [Analysis of CPHS Requirements
    ]
    ]
    ]
    [\textsection \ref{sec:s_threats} Security Threats at different Layers
    [\textsection \ref{subsec:tlt} IoT Devices]
    [\textsection \ref{subsec:clt} IoT-based Communication]
    [\textsection \ref{subsec:alt} IoT Applications]
    [\textsection \ref{subsec:ai} AI-based Security Approaches]
    ]
    %[Research Gaps]
    [\textsection \ref{sec:psolution} Proposed Solution
    [\textsection \ref{subsec:wcphs} Workflow of CPHS]
    [\textsection \ref{subsec:bcphs} Features of CPHS]
    ]
    [\textsection \ref{sec:conclusion} Conclusion]
  ]
\end{forest}
    \caption{Logical Structure of the Paper Organization}
    \label{fig:paperstructure}
\end{figure}

%%%%%%%%%%%%%%%%%%%
\begin{table*}[ht]
\caption{Developments of Healthcare (aka Medicine) 1.0 - 5.0}
\centering
\begin{tabular}{|>{\centering\arraybackslash}m{1.2cm}| >{\centering\arraybackslash}m{7.2cm}| >{\centering\arraybackslash}m{2cm} | >{\centering\arraybackslash}m{5cm}|}
%\begin{tabular}{|l| l |l| l|}
\hline
\hline
\rowcolor{gray!20}
\textbf{Healthcare} & \textbf{Developments} & \textbf{Era}	& \textbf{Focus (Medicine/Healthcare) } \\ [0.5ex]
\hline
\hline

 1.0 & \specialcelll{Natural substances (herbs)\\ Sanitation, germ theory, vaccination}	 & Production & Public health/ Evidence-based treatment \\ 
\hline

  2.0 & \specialcelll{Antibiotics, X-rays\\ Big hospitals, professionalism, specialists} & Industrialising & Mass production/Value chain  \\
\hline

 3.0 &\specialcelll{Image recognition, evidence-based, microtechnology\\ Computer imaging and evidence-based}  & Automation & Information technology/Operating model  \\
\hline

 4.0 &\specialcelll{Communication + Microtechnologies\\ Artificial intelligence, precision, telemedicine} & Digitalisation & Smart health/Business model \\ 
\hline

 5.0 &\specialcelll{Connectedness + Integrity\\Customer models, lifelong partnership, digital wellness}  & Personalization & Personalized medicine/Customer model\\ [1ex]
\hline
\end{tabular}

\label{table:hce}
\end{table*}
%%%%%%%%%%%%%%%%%%%
\section{Transformation of Healthcare Technology}\label{sec:thc}
%\section{\textbf{Healthcare 5.0}}
The history of healthcare systems is deeply connected with the history of medicine. As shown in Figure~\ref{fig:h1-5} medicine's history starts with Medicine 1.0, where it was dependent on highly qualified doctors, and most of the medication was based on natural substances like herbs. These were the days of Healthcare 1.0, where intelligent public health approaches are used to solve significant health problems~\cite{Chen2019}. The discovery of antibiotics and the use of X-rays for diagnostic purposes has introduced the concept of big hospitals and specialization in diagnostic techniques and is given the name Healthcare 2.0. Healthcare 2.0 was the era of industrialization and its primary structure was mass production (aka Medicine 2.0). Later surgery was benefited a lot with the new advancements in electronics and micro-technology. Evidence-based medicine, the introduction of surgical robots, navigation surgery, and image recognition had given birth to Medicine 3.0 and was the basis for Healthcare 3.0. With the advancements in information systems and technologies, intelligent devices with smart microelectronics, fast transmission and network technologies are working together in healthcare systems and are playing a significant role towards the better quality lifestyle and improved healthcare services which ultimately lead towards a healthier life, which is called Medicine 4.0 (aka smart-health). Some of the examples of Medicine 4.0 are personalized chemotherapy, personalized telematic therapy, intelligent implants for cancer therapy, and intelligent occlusal splint~\cite{wolf}. 
 The focus of Healthcare 4.0 was to develop business models for healthcare systems. Smart wearables with integrated sensors help to collect, monitor, and diagnose diseases from the patient's data using artificial intelligence (AI)/machine learning (ML) techniques. However, IoT devices with artificial intelligence cannot be considered as a solution to the limitations (scalability, security, privacy, and reliability) in fourth-generation healthcare systems due to the following major challenges at the communication layer,
\begin{itemize}
    \item seamless data transmission rate with minimum or no data loss,
    \item traffic-free transmission channels, 
    \item cost-effective, 
    \item no time data retrieval and 
    \item machine to machine (M2M) or device to device (D2D) communication.
\end{itemize}
The emergence of advanced communication technologies like 5G has resolved the major challenges at the communication layer that resulted in connectedness and integration, and is the start of Medicine 5.0 (aka Healthcare 5.0 ). The main focus of Healthcare 5.0 is personalization by developing customer models, digital wellness, and considering the wellbeing of not only patients but individuals. 
The developments of  healthcare/medicine 1.0 - 5.0 are shown in Table~\ref{table:hce}. Other similar initiatives around the globe are Industrie 4.0 (Germany and EU), Cyber-physical systems (CPS), Society 5.0 (Japan), and Cyber social systems \cite{wolf}.

Based on the introduction to Healthcare 5.0,  in the following subsection, we define personalized healthcare.
\subsection{Personalized Healthcare}\label{subsec:ph}
 According to the World Health Organisation (WHO), healthcare is defined as the maintenance or improvement of a complete state of physical, mental, and social well-being
    whose main purpose is to enhance the quality of life and to extend the life expectancy of patients. This can be achieved by enhancing their personalized health conditions.

\begin{figure}[ht]
    \centering 
    \includegraphics[scale=0.485]{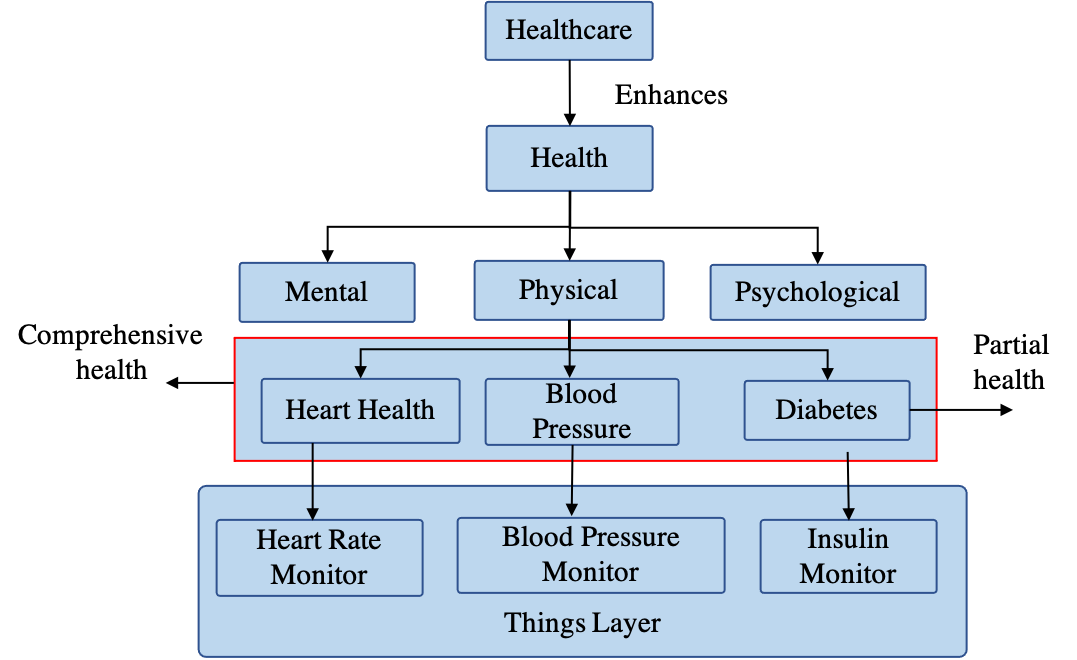}
    \caption{Partial vs Comprehensive Health}
    \label{fig:chealth}
\end{figure}

Current approaches consider personalized healthcare as a service that tailors medical treatment to individual patients through the identification of common features, including their genetics, inheritance, and lifestyle~\cite{zhang2019}. However, this consideration helps to personalize a single health condition with specific causes (e.g., genetics, inheritance) but is unable to personalize the overall health of a patient with multiple health conditions which requires an understanding of the biological relationship among different health conditions. Furthermore, particularities of personalized healthcare are different for different patients as everyone has different health condition determinants, namely, biological, physiological, and psychological characteristics (e.g., immunity, sensitivity against variable phenomena) that play a vital role in maintaining someone's personalized health, which implies that patient's health condition parameters will also be unique.

% \begin{itemize}
%     \item biological,
%     \item physiological, and 
%     \item psychological characteristics 
%     \\
%     (e.g., immunity, sensitivity against variable phenomena) that play a vital role in maintaining someone's personalized health, which implies that patient's health condition parameters will also be unique.
% \end{itemize}
To this end, we view personalized healthcare as a service that provides determinant-based optimization of multiple health conditions of a patient (which we call comprehensive healthcare service - CPHS). Determinant-based optimization aims to improve the long-term overall health of a patient by reducing side-effects of different health conditions of the patient (e.g., heart condition, blood pressure condition, insulin condition), while considering the relation among determinants. Healthcare 5.0 can make it possible to establish a comprehensive personalized healthcare service (CPHS) for patients and improve their long-term health. Figure~\ref{fig:chealth} sketches the difference between partial and comprehensive health.

 In the following subsection, we used an example use case scenario based on the CPHS.
\vspace{-0.2cm}
\subsubsection{\textbf{Example Use Case Scenario}}\label{sec:ucs}
\begin{figure}[ht]
    \centering 
    \includegraphics[scale=0.5]{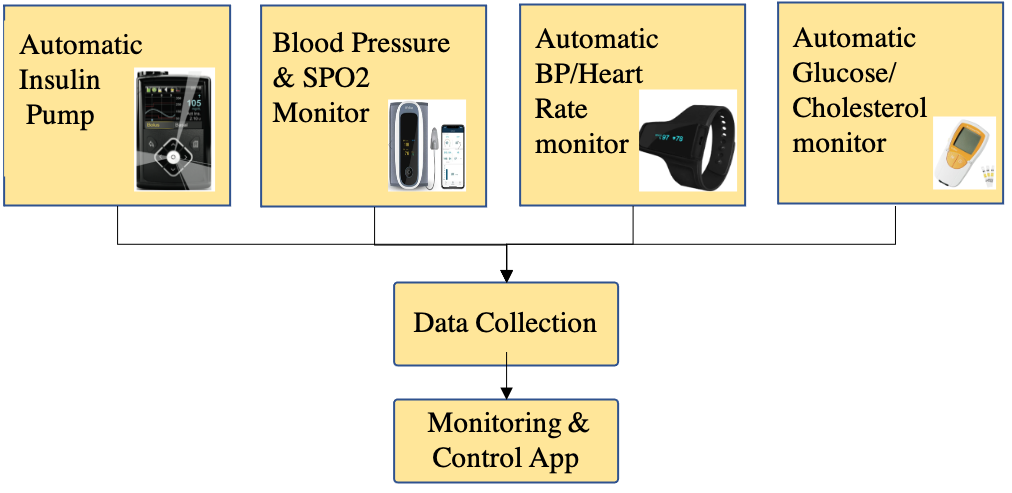}
    \caption{Example Use Case Scenario}
    \label{fig:usecase}
\end{figure}

Figure~\ref{fig:usecase} shows different health monitoring devices (like automatic blood pressure monitor, automatic blood glucose and blood sugar monitor, automatic heart rate monitor, and automatic insulin pump) that monitor different health conditions (like blood pressure, diabetes, heart, and cholesterol) of a patient. The monitoring devices share underlying health condition data collected during the data collection phase, for further processing. Later, the collected data is used by other applications to monitor and control different health conditions when required. 

Currently, the applications monitor and control every monitoring device separately mainly because each device is manufactured for control and monitoring of a dedicated health condition whose details are in Table~\ref{table:hmd}. However, this is not practical because, in reality, various health conditions are interdependent. For instance, a certain health condition (e.g., high blood pressure) could be a consequence of a side-effect of another specific health condition (e.g., high cholesterol). Therefore, we describe a use case scenario that aims to provide optimized and rigorous personalized health conditions of a patient through analyzing intra-dependencies among different health conditions of the patient. To this end, we consider the following two variations of the dependencies:

\begin{itemize}
    \item One-to-Many (1--*), where a health condition may affect none or many other health conditions, e.g., common cold, fever. 
    \item Many-to-Many (+ -- +), where various health conditions affect various other health conditions, e.g., arthritis, diabetes and epilepsy.
\end{itemize}
As sketched in Figure~\ref{fig:usecase}, in our example use case, the data is first collected from different monitoring devices and then used to analyze dependencies among the monitored health conditions as follows:
\begin{itemize}
    \item No-dependencies - each health condition data is used to determine if it appropriately characterizes the underlying health condition. For instance, common cold or fever has no long-term effects on a person's blood glucose/ cholesterol levels.
    \item Dependencies - data from different health conditions determine side-effects of various other health conditions. Which results in getting other health conditions. For instance, medical conditions interact with one another in the following ways:
    \begin{itemize}
        \item One medical condition might make another worse like someone with arthritis who finds difficulty in exercise got a heart or lung problem.
        \item One medical condition could start other serious medical conditions like diabetes contributes to cardiovascular disease, high blood pressure that leads to a heart attack or stroke.
        \item May clash with each other, making one ineffective or producing side effects.
         \item Treating one medical condition might make the symptoms of another worse or may produce a new or previously hidden health condition \cite{yan2020}.
    \end{itemize}

\end{itemize}
Based on the identified dependent health conditions, their current treatment, and other associated patient's routine (e.g., exercise and food), control applications determine a target personalized health condition by optimizing the involved health conditions. To this end, the application issues appropriate control alarms manipulated automatically or semi-automatically to establish the health condition.

\begin{table*}[t]
\centering
\caption{Health Monitoring Devices}
\begin{tabular}{|>{\centering\arraybackslash}m{3cm}| >{\centering\arraybackslash}m{4cm}| >{\centering\arraybackslash}m{4cm}| >{\centering\arraybackslash}m{2cm}| >{\centering\arraybackslash}m{2cm}| >{\centering\arraybackslash}m{2cm}|}
 \hline
 \hline
 \rowcolor{gray!20}
 \textbf{Things} & \textbf{Product Name} & \textbf{Vendors} & \textbf{Communication} & \textbf{Accuracy} \\
 \hline
 \hline
Blood Pressure Monitor & Viatom Blood Pressure and SP02 Monitor & Viatom & Bluetooth & $\pm$3mmHg  \\\hline

Automatic Insulin Pump & MiniMedTM 640G & MiniMedTM & Radio & $\pm$5\% to $\pm$20\%  \\\hline

Blood Cholesterol Monitor & Accutrend Plus & Roche & Biosensor & -1.29 - 0.98 \\\hline

Heart Rate Monitor & Viatom Sleep Oximeter With Vibration & Viatom & Ring Sensor & 2 bpm or $\pm$2\%  \\\hline
\end{tabular}
\label{table:hmd}
\end{table*}
%%%%%%%%%%%%%%%%%%%%%%%%%%%%%%%%%%%
%%%%%%%%%%%%%%%%%%%%%%%%%%%%%

In the following subsection, we define the key requirements for establishing CPHS.
\subsection{Key Requirements of CPHS}\label{subsec:keyreqs}
Healthcare 5.0 is the emergence of digital wellness~\cite{singh2020} and it aims to provide personalized healthcare services by integrating cyber (e.g., monitoring/control) and
physical (e.g., biological) processes/components of the service. Their integration helps to develop personalized healthcare services that are autonomous supporting automatic control of healthcare services by remotely monitoring the health conditions. Since these services are part of the critical application domain, therefore such services should meet certain critical requirements~\cite{Somayeh2019}~for service operations that are given below:
\begin{enumerate}
\item\textbf{Reliability Requirements:} need to address the following key attributes: 
    \begin{itemize}
        \item \textbf{Confidentiality:} ensures that only authorized users and devices can use medical data.
        \item \textbf{Integrity:} ensures the data completeness and accuracy in the entire communication to avoid wrong diagnosis and prescription.
        \item \textbf{Availability:} ensures the accessibility of medical data and devices to the authorized users at the time of need without any failure. 
        \item \textbf{Privacy:} ensures that the privacy policies must be followed and any sensitive and private data must not be disclosed or shared without consent.
        % \item \textbf{Non-repudiation:} ensures that no one can deny the completed actions.
        \item \textbf{Data Freshness:} ensures that the physician or the system will get the recent data to rightly diagnose and prescribe and supports real-time response. For example the fresh data for blood glucose level in fasting to set the right amount of insulin. 
        \item\textbf{Security:} ensures the three fundamental security elements, i.e., “privacy", "integrity” and “accessibility".
    \end{itemize}
\item\textbf{Resilience Requirements:} IoT-based healthcare system includes the following~\cite{Somayeh2019} requirements:
        \begin{itemize}
            \item \textbf{Protection,} involves:
            \begin{itemize}
                \item \textbf{Self-healing:} is the process in  which the system identifies the failure of medical devices and is able to recover the software and hardware automatically without any data loss.
               
           \item \textbf{Self-optimizing:} is the process in which the system will improve its performance, quality of service and can automatically optimize resource consumption like energy, memory, etc.
                \item \textbf{Self-protecting:} is the process in which the system automatically protects itself against harmful attacks and generates alarms in case of any suspicious/failure event.
                \item \textbf{Self-configuring:} is the process in which the system automatically installs, configures, and integrates itself to eliminate system's flaws.
            \end{itemize}
            \item \textbf{Safety,} ensures the safety of data and devices (i.e., the safety of the patient as well as of the involved cyber and physical components). 
        \end{itemize}
\item\textbf{Personalized Requirements:} These are the characteristics that are personalized to every patient and are a factor of response to the patient's health condition, immune system, and other underlying associated health conditions. Here are few such requirements: 
    \begin{itemize}
        \item \textbf{Deep-analysis:} ensures that each health condition of a patient is determined by analyzing various aspects of the condition including, mechanical and biological processes of the condition monitoring. The immune system of different patients may respond differently against the same disease. Furthermore, different medical devices used for monitoring health conditions may have different characteristics, e.g., error rates.
        \item \textbf{Coordinated-healthcare:} ensures that the personalized health condition is resolved as a comprehensive health condition, and is possible through deep-analysis of each health condition of the patient and by identifying various (mechanical and biological) dependencies among different health conditions of the patient. This helps to identify the strictly personalized health condition of the patient.
    \end{itemize}
\end{enumerate}

\section{Reference Architecture for IoT based Healthcare System}\label{sec:r_architecture}

\begin{figure*}[ht]
    \centering
    \includegraphics[scale=0.65]{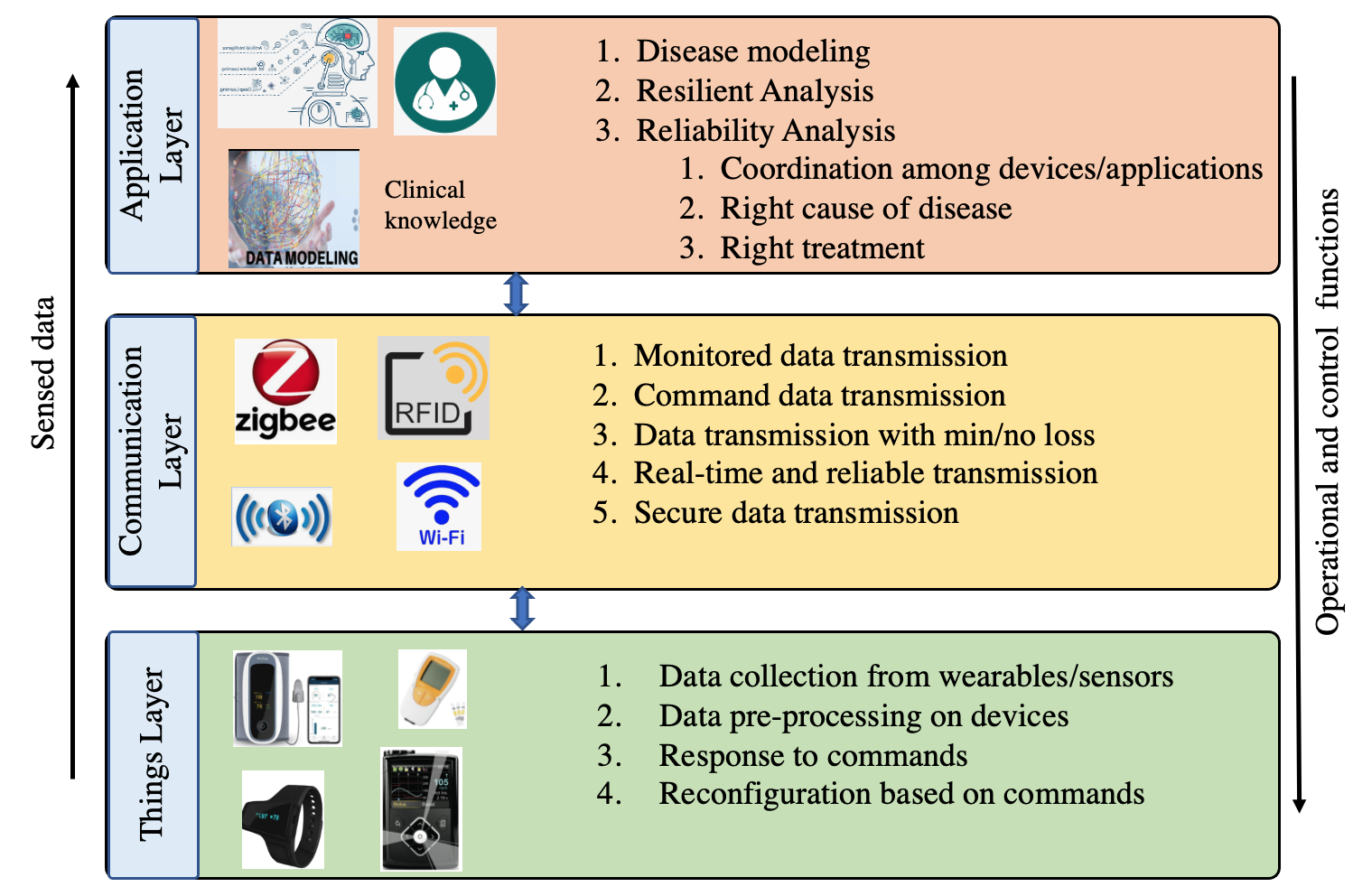}
    \caption{HIoT System Architecture}
    \label{fig:hiot system}
\end{figure*}

The word Internet of Things (IoT) is a combination of two words Internet and Things. Every object capable of connecting to the internet is considered as Thing, which includes smart devices, sensors, and context-aware objects that can communicate and are pervasive. Considering IoT as fully pervasive, IoT devices must possess the following three basic qualities:

\begin{itemize}
    \item \textbf{Ability to Sense}: IoT devices must carry the ability to sense and combine the sensed data. For example, in the healthcare field, blood glucose, heart rate, body temperature, cholesterol level are sensed and aggregated using different specific biosensors. The data aggregation is autonomous.
    \item \textbf{Communicable}: After sensing and aggregating, IoT devices should be able to transmit data to the various desired data centers by using different communication mediums (i.e., Wireless Technologies, Mobile networks (3G, 4G, 5G), wireless local area networks (WLAN), wireless sensor networks (WSNs), and Mobile Adhoc Networks (MANET), etc.).  
    \item \textbf{Actionable}: The aggregated data alone doesn't make any sense so, in the case of responsive devices IoT devices must be able to process the data in order to take some action. For example, if a patient’s pulse rate or blood pressure level is sensed and communicated, alone this data is useless unless some processing technology is applied and the system finds that if it exceeds/falls behind from normal range. IoT systems also need to be capable to send automatic alerts to healthcare providers for further interventions.
\end{itemize}
We consider HIoT architecture based on three tiers. This three-tier structure consists of the Things layer followed by the communication and application layer. The description of the HIoT system and its operational and control functions is shown in Figure~\ref{fig:hiot system}

In the following sub-sections, we analyzed the three layers of the IoT systems. In each sub-section, we 
\begin{itemize}
    \item discussed a general overview of the layer,
    \item identified key challenges to establish the identified requirements of HIoT (e.g., reliability, resilience, and personalization),
    \item analyzed current approaches (AI/non-AI-based) and to address the above challenges, and
    \item identified the challenges as research gaps that have not been addressed adequately.
\end{itemize}

\subsection{Things Layer}\label{subsec:tl}
\begin{figure*}[ht]
    \centering
    \includegraphics[scale=0.55]{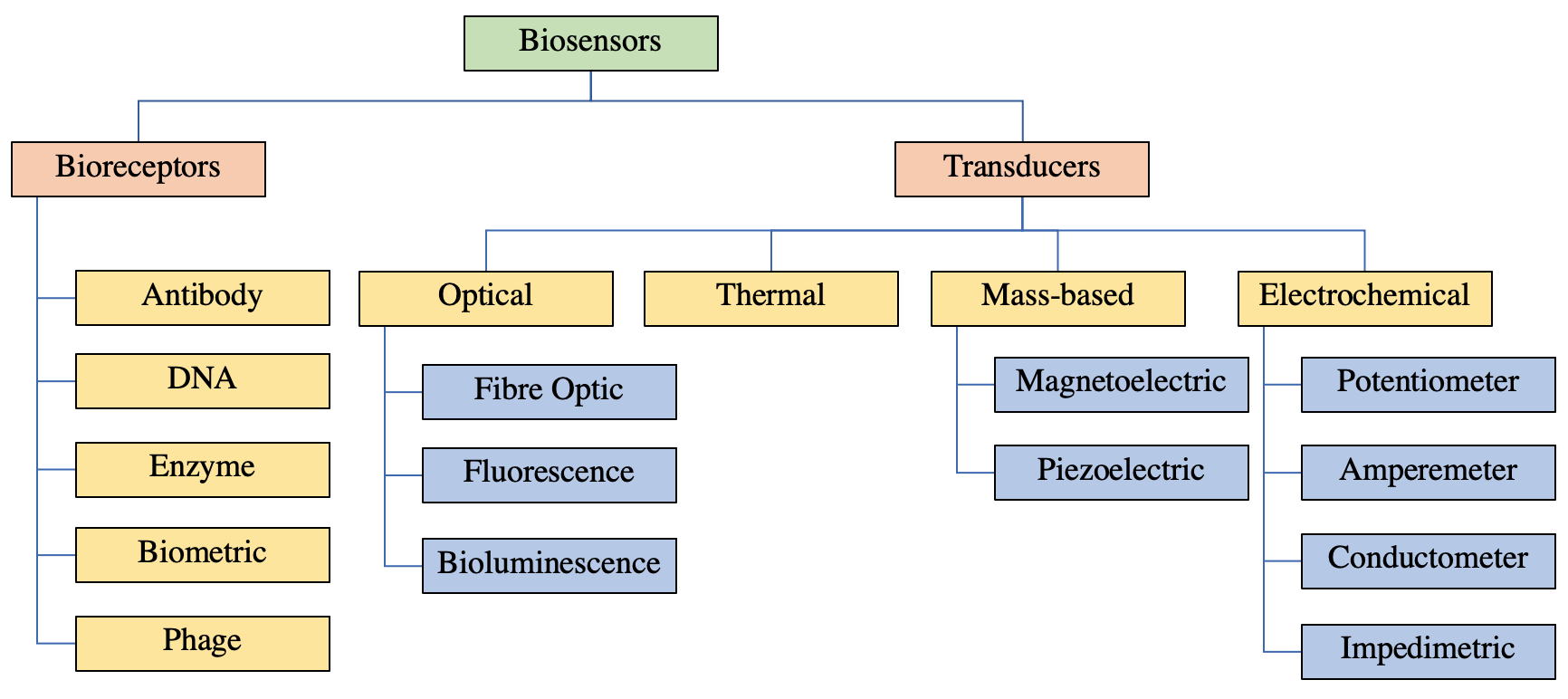}
    \caption{Typical Nomenclature of Biosensors}
    \label{fig:lbiosensors}
\end{figure*}

IoT devices are seamlessly integrating into healthcare and have reformed the healthcare industry with their multiple healthcare monitoring devices and applications. IoT-based solutions like smart sensors, wearable devices, and smart health monitoring systems play significant roles in developing healthcare systems (smart hospitals, mobile healthcare (mHealth)), and the healthcare industry. 
Things layer~\cite{Qadri2020} in HIoT, are sensing systems and devices that record the different values observed from different sources (i.e., devices monitoring patient's different body parts or health conditions) based on the application. Things layer consists of different devices and Things, that include:
\begin{itemize}
    \item \textbf{Sensors,} whose aim is to sense/detect the physical quantities like temperature, pressure, smoke, light, etc., and then convert it into desired output in the form of the electrical signal to measure the applied physical quantity. Sensors are further categorized into:
    \begin{itemize}
        \item \textbf{Biosensors} are analytical devices used in detecting the presence or concentration of a biological analyte, such as a biological structure, molecule, or a microorganism in the living body. Biosensors usually consist of three parts: 
        \begin{itemize}
            \item\textbf{Chemical Component}, whose job is to recognize the analyte and produce a signal,
            \item\textbf{Transducer}, that converts this signal to an electrical signal and 
            \item\textbf{Reader Device} where this signal is displayable.
        \end{itemize}
        Figure~\ref{fig:wpbiosensors} sketches biosensor operation and its main components.
        
    \begin{figure*}[ht]
    \centering
    \includegraphics[scale=0.50]{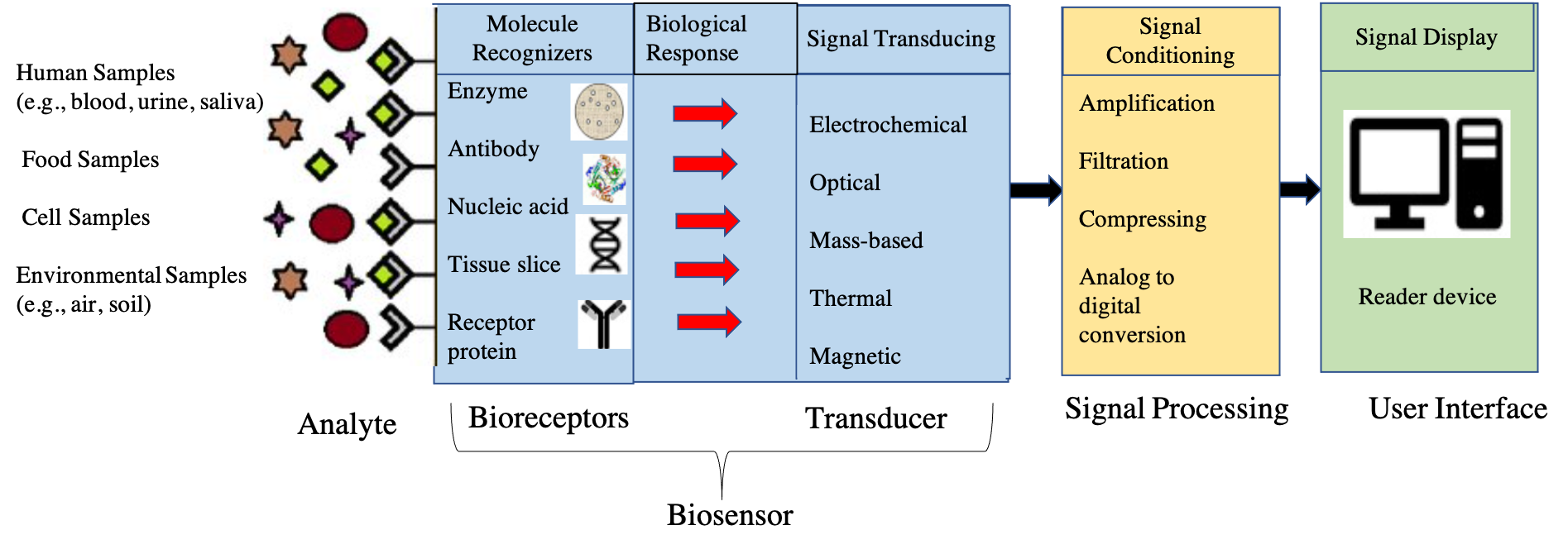}
    \caption{Workflow and Components of Biosensors}
    \label{fig:wpbiosensors}
\end{figure*}

Biosensors play a vital role in the medical field, health condition monitoring, clinical analysis of different health conditions, and diagnostic applications. They are advantageous over the lab equipment due to many reasons. Some of them include small size, low cost, quick results, reusable, and avoiding contamination.
% are as follows:
% \begin{itemize}
%     \item Small size
%     \item Low cost
%     \item Quick results
%     \item Reusable 
%     \item Avoid contamination
% \end{itemize}
 A comprehensive list of biosensors based on their working mechanism is shown in Figure~\ref{fig:lbiosensors}. Further different types of biosensors, their details with their working principles and the diagnosis are shown in Table~\ref{tab:biosensors}. These sensors aims at detecting disease-specific data from a patient's body.
        \item \textbf{General-Purpose Sensors} like temperature sensor, pressure sensor, proximity sensor, accelerometer and gyroscope sensor, infrared (IR) sensor, optical sensor, gas sensor, smoke sensor, and many more. These sensors aims at detecting various environmental conditions required for a patient's health.
    \end{itemize}
    \item \textbf{Wearable Monitoring Devices} are the sensors that are worn anywhere on the human body in the form of some object that allows continuous or intermittent monitoring of physiological signals. They play an important role in healthcare, industrial process control, online control, offline control (distant and local), military applications, and continuous environmental monitoring. Some examples are ring sensor, smart shirt, heart rate monitor, blood pressure monitor, blood sugar monitor, stress monitor, smartwatches, textiles, glasses, patches, etc. These wearable biosensors are used for different applications e.g.,
    \begin{enumerate}
      
\item \textbf{Ring Sensors} are used for:
    \begin{itemize}
   \item  Wireless supervision of people during hazardous operations in the military during fire fighting.
   \item In an overcrowded emergency department.
    \item In cardiovascular disease for monitoring the hypertension. 
    \item For chronic surveillance of abnormal heart failure.
    \end{itemize}  
    The main advantages of using wearable monitoring devices are:
    \begin{itemize}
   \item Continuous monitoring,
   \item Detection of transient phenomena,
   \item Promote further diagnostic and therapeutic measures,
   \item Easy to use, flexible, and 
   \item Reduces hospitalization fee.
    \end{itemize}  
Considering their use Wearable devices are expected to reach a market value of \euro 24.4 billion globally by 2023. There have been various developments in wearable devices. In~\cite{Wang2018} the authors presented a novel ring sensor for the continuous measurement of sympathetic nervous system activities. They successfully tested it on different age groups and got accurate results compared to the state-of-the-art open-source devices.
The technology used by wearable gadgets is with low processing capabilities. Some examples are Google Glass and Smartwatch. Mobile devices are used as the computation source to increase the impact of these wearables technology. However, in terms of device security and privacy, this wearable technology is not as mature as it needs to be. These devices are vulnerable to attacks due to the security and privacy issues where due to the low processing power,  the developers can't use complicated security mechanisms and algorithms~\cite{Ching2016}. Some of the disadvantages of wearable monitoring devices are
\begin{itemize}
    \item Initial cost is high
    \item Limited number of physiological parameters for  monitoring
    \item battery life is short\\
   The battery life of wearable monitoring devices is addressed In~\cite{Hooshmand2017} where authors used the original dictionary-based technique where battery life for these devices used in healthcare is achieved by compressing the biosignals, and this compression is achieved by building and maintaining a dictionary at run time.
\end{itemize}
%%%%%%%%%%%%%%%%%%%%%%%%%%%%%%
\begin{table*}[ht]
\centering
\caption{List of Biosensors with Types, Principle and Diagnosis~\cite{YIN2016}}
\begin{tabular}{|>{\centering\arraybackslash}m{3cm}|>{\centering\arraybackslash}m{3.5cm}|>{\centering\arraybackslash}m{3.5cm}|>{\centering\arraybackslash}m{3cm}|>{\centering\arraybackslash}m{1.5cm}|}
 \hline
 \hline
 \rowcolor{gray!20}
 \textbf{Sensor Type}  &  \textbf{Sensor Name} & \textbf{Diagnosis} & \textbf{Principle}  & \textbf{Portability}  \\ 
 \hline
 \hline
Electrochemical~\cite{Clark1962} &	Glucose Oxidase Electrode Based Biosensor &	Analysis of Glucose in Biological Sample &	Electrochemistry using Oxidation &	No  \\ 
\hline
\specialcelll{Optical/Visual/\\Fluorescence~\cite{Dhankhar2020}}	& Stilbene-420 	& eye glasses, biosensing & Optical/Visual/Fluorescence & No \\
\hline
Microbial~\cite{Sun2015} &	Microbial Fuel Cell-based Biosensors &	Energy production and environmental sensing &	Optical	& Yes	 \\ 
\hline
Electromagnetic~\cite{wang2020} & Quartz Crystal Biosensor & TNT Explosive Detection & Electromagnetic	& No \\ 
\hline
Electrochemical~\cite{Franco2020} & carbon-based screen-printed electrodes & water quality monitoring &	Electrochemistry &	No \\ 
\hline
Enzyme-based~\cite{Singh2019} &	Glucose Oxidase Biosensor for Diabetes & Diabetes &	oxidation of Beta-D-glucose & Yes  \\ 
\hline
Amperometric Biosensors~\cite{Rev2020} & Nitrite Biosensors &	Environmental Analysis of Nitrate, Nitrite, and Sulfate &	Enzymatic &	Yes \\ 
\hline
DNA Biosensors~\cite{fani2020} & Electrochemical DNA Biosensor  & Detection of Human T-Lymphotropic Virus-1&differential pulse voltammetry (DPV) & Yes \\ 
\hline
Enzyme-based~\cite{Rawat2020} & Thiol Biosensor & Highly Selective Detection of Glutathione	& Electrochemical &	Yes	 \\ 
\hline
Enzyme-based~\cite{Malik2019} &	Implantable Glucose Biosensor &	Continuous Glucose Monitoring & RF sensor & Yes  \\ 
\hline
Enzyme-based~\cite{Kumar2019} &	Cholesterol Biosensor &	Cholesterol & Single-Mode Fiber (SMF) and a Hollow Core Fiber (HCF) &	Yes	 \\
\hline
Enzyme-based~\cite{Schroder2017}&	Nitric oxide Biosensor & Detecting NO Gas  & Conventional Wire Bonding Technology & Yes  \\ 
\hline
Enzyme-based~\cite{Yang2020} &	Superoxide Anion Radical Biosensor & Evaluating the photosensitivity of colorful wastewater & 	Chemical Reaction	& Yes  \\ 

\hline
\end{tabular}
\label{tab:biosensors}
\end{table*}

%%%%%%%%%%%%%%%%%%%%%%%%%%%%%%%%%%%%%%%%%%%%%%%

    \item\textbf{Medically Implanted Devices/Sensors} are used in the human body in the following ways:
    \begin{itemize}
        \item For replacing a missing biological structure,
        \item for supporting a damaged biological structure, and
        \item for enhancing an existing biological structure
    \end{itemize}
    using biomedical material tissues, active implant electronics, and transplanted biomedical tissues. Some of the examples of medically implanted devices are cochlear implants, automatic insulin pumps, pacemakers for cardiovascular diseases, retinal implants, artificial heart, cardiopulmonary bypass, and many more.
    \end{enumerate}
\end{itemize}

The above-mentioned sensors and devices collect data from the patient's body and in some cases after small processing transfer this data to the data center where this big data is analyzed. In some responsive sensors, the devices monitor the related body parameter and after processing perform the operation for the body like in smart automatic insulin pumps, the sensors check the blood sugar level of the patient in specific time intervals and based on the disease modeling of the device performs operations and automatically inject insulin into the body of the patient. We have highlighted key challenges of the Things layer that hinders key requirements of HIoT as identified in the Sub-section~\ref{subsec:keyreqs}. Furthermore, we have analyzed current approaches (AI/non-AI-based) to address the challenges. Additionally, we have also identified the challenges as research gaps that have not been addressed adequately.
In the following, we have discussed the above for each requirement, namely reliability, resilience, and personalization, in order below.

\begin{enumerate}
 \item \textbf{Reliability:}
 Reliability is the probability of a component or a  system reaching specific performance standards and producing the desired output in certain environmental conditions for a specific time duration~\cite{pokorni2019}. In other words for IoT devices, reliability can be defined by considering the following three requirements~\cite{kouicem2018}.
 \begin{itemize}
     \item Perform a required function (i.e., Things are performing their required functions as expected, which is mainly sensing of various parameters).
     \item Perform the function under stated conditions (i.e., Things are functioning in the desired setting/configuration and environment)
     \item Operation for a specific time (i.e., Things are automatically functioning at a configured time for a specific time)
 \end{itemize}
 Reliability is one of the main challenges in sensor technology and in the more sensitive systems like healthcare reliability is of vital importance. The major challenges that hinder reliability of Things layer include:
  \begin{itemize}
     \item  Constrained nature of the Things that contains battery, memory, and computational capacity~\cite{al2015,kouicem2018}.
    %  the following~\cite{al2015}
    %  \begin{itemize}
    %      \item battery
    %      \item memory and 
    %      \item computational capacity\cite{kouicem2018}
    %  \end{itemize}
     \item Conditions of the operational environment (e.g., heat, freezing temperatures, mechanical or spontaneous wear, vibration, and moisture)\cite{rayes2017}.
     \item Tendency of sensors to "fail-dirty" (a phenomenon in which the sensor continuous to send fallacious information after some failure)\cite{kim2016, karkouch2016}.
 \end{itemize}
In~\cite{Mr2019} the authors presented a smart and secure framework for the hospital environment using the Internet of Things (IoT) and Artificial Intelligence (AI). The presented system solved information system, treatment, diagnosis, patient monitoring, and record maintenance problems effectively. According to~\cite{alam2018survey,lin2017survey,choudhury2019} AI-based algorithms are able to process huge amounts of data that are collected from different smart IoT devices within split seconds and predict the result, these predicted results along with the electronic health reports (ERHs) can be further analyzed by medical partitions. AI-based algorithms are able to think and detect illness faster with better accuracy and can assist the practitioners about the possible illnesses that can cause serious health problems in the future and can suggest possible treatment and medications for these diseases.
 In~\cite{le2018} artificial neural networks (ANN) are used to analyze urine and blood samples, as well as track glucose levels in diabetics, determine ion levels in fluids, and detect various pathological conditions.

 Above mentioned challenges are addressed in different ways. For instance,~\cite{vergutz2020,alam2020} proposed solutions to improve the reliability of sensors by making sensors capable of obtaining and providing reliable and accurate data. Evaluation of sensor reliability involves statistical or probabilistic data that multiply the reliability estimate. According to~\cite{lay2015} the development of "smart sensors" can improve the reliability of the sensors.
 Current devices operating at the Things layer are not rigorously reliable to support real-time and critical operations of HIoT. For instance, some devices suffer from discrepancies by design~\cite{Minimed2016}, e.g., the error rate of blood pressure monitor. Similarly, the others may be faulty either due to their maligned software (e.g., as a result of security attacks)
 ~\cite{lv2020, ande2020} or for some mechanical failure (e.g., as a result of mishandling of the device by an elderly patient). Moreover, such devices also fail to operate reliably due to their limited resources, e.g., battery life of sensors, signal strength, memory, and computational capacity~\cite{lopez2015, yang2017, kouicem2018}.

     \item\textbf{Resilience:} 
    Resilience is the capability of systems to continue their normal operations and vigorous response in case of any unexpected or unpredicted situations. The importance of resilience multiplies in sensitive systems like healthcare where the continuity of the system is desirable from diagnosis to treatment and follow-up. Gaps in the continuity of care threaten a patient’s well-being and cause adverse events. According to \cite{rajamaki2016}, resilience is two-fold; a system must be sound enough against attacks (that is in the first place be able to obstruct most attacks) and it must be able to go back to a safe state after any occurred attack. Abnormal operations at the Things layer occur due to several reasons. Some of them are as under:
    \begin{itemize}
        \item Hardware failure (e.g. sensors failure,  failure of the control unit of IoT due to hardware and/or software, etc.)\cite{khalil2020,letafatimehdi2020}.
        \item Failure of cloud services~\cite{sharma2016,Yanovsky2016} that may include failure of software, hardware, scheduling, service, power, and denser system packaging. 
        % \begin{itemize}
        %     \item software failure
        %     \item hardware failure
        %     \item scheduling
        %     \item service failure
        %     \item power outage
        %     \item denser system packaging
        % \end{itemize}
    \end{itemize}
    %%%%%%%%%

    There have been some efforts to deal with the different issues of IoT. In~\cite{Strielkina2018} the authors focus on the IoT healthcare infrastructures, failures of components (hardware, software), and complete systems. The reasons for their failure may vary depending upon the healthcare IoT infrastructure.  
  According to~\cite{Tsigkanos2019} important challenges that hinder the safety and failure-free execution of IoT are as under:
  \begin{itemize}
      \item Capabilities and technical specifications of IoT's. 
      \item Geographically dispersed deployment.
      \item Lack of security policies.
      \item Diverse attacks on IoT, and management of security patches.
  \end{itemize}
  So far, various approaches have improved the resilience of healthcare IoTs~\cite{islam2015, laplante2018, kharchenko2016, Strielkina2017, Strielkina2018}. In~\cite{islam2015} authors proposed an intelligent collaborative security model to minimize the security risks, but the possible failures of IoT components and systems are not handled and their influence on the infrastructure is still unsolvable. In~\cite{laplante2018} the authors used traditional techniques of safety and violence for describing failure and misuse cases.The works in \cite{Strielkina2017, Strielkina2018} proposed the Markov Queuing approach for taking into account the safety and security issues of healthcare IoT infrastructure. In~\cite{Strielkina2018} the author proposed the Markov model that considers possible failures of healthcare IoT systems as well as recovery procedures.
  
  In~\cite{letafatimehdi2020} the authors proposed a lightweight security scheme for ensuring transmission resilience and information confidentiality in the Internet-of-Things (IoT) communication. In their proposed system a single-antenna transmitter communicates with a half-duplex single-antenna receiver in the presence of a sophisticated multiple-antenna-aided passive eavesdropper and a multiple-antenna-assisted hostile jammer (HJ). The proposed system is demonstrated using simulations. However, some technical challenges like network synchronization and interference cancellation need to be addressed in experimental validation. In~\cite{garg2020} the authors designed a novel blockchain-enabled authentication key agreement protocol BAKMP-IoMT for the internet of medical things (IoMT) environment that provides secure key management between implantable medical devices and personal servers, and also between personal servers and cloud servers. The proposed design stores data using blockchain and this data can be accessed by the users in a secure way using cloud service. The developed tool provides resilience against different types of possible attacks.
  
  Current devices operating at the Things layer are not resilient, i.e., they fail to recover in case of any failure or physical damage. AI and blockchain-based technologies provide data transmission and storage resilience. The devices support limited hard resilience. They support resilience either by restarting the device in case of software failure or by replacing the device in case of mechanical failure/damage.
  
  \item \textbf{Personalizing:}
  With the exponential growth of commercially available wearable devices, the concept of IoT-based personalization healthcare services is well-established and well-liked. These healthcare services use a set of interconnected devices and create an IoT-based healthcare network that is capable of observing and recording multiple types of health related data. After collecting data,  different healthcare activities are performed such as monitoring, diagnosis, treatment, and remote surgeries~\cite{khan2020, datta2015,kau2015}.
  In~\cite{ZhangHe2020} the authors developed smart socks using textile-based triboelectric nanogenerator (TENG). The authors developed Artificial intelligence (AI)-based algorithms,  for gait analysis and provide detailed information of multiple triboelectric socks users.  It helps to identify various health conditions and activities of users for smart home and healthcare applications. 
  In~\cite{HALEEM2021} the authors tried to access the capabilities of AI for predicting congestive heart failure for the COVID-19 patients. AI can help to provide advanced cardiac treatment, and to analyze-/measure the functioning of the human heart.
  
  According to~\cite{Jagadeeswari2018, khan2020} the major challenges for personalised healthcare are: 
  \begin{itemize}
      \item Data privacy and security,
      \item handling big data (as some healthcare sensors/wearables are working 24/7),
      \item healthcare application are inadequate and may suffer: 
      \begin{itemize}
          \item Low quality of data deliverance.
          \item Performance analysis.
          \item Data storage.
          \item Privacy prevention, and standardization.
      \end{itemize}
  \end{itemize}
The authors used different approaches to address the above-mentioned challenges like cryptographic techniques for attaining data privacy and security,
cloud storage system to handle huge healthcare data and smartphone-supported self-monitoring sensors for better quality performance.
     Current devices operating at the Things layer provide limited personalized healthcare services. For instance, they can be configured to monitor a specific health condition (e.g., chronically ill patients, active and assisted living, dementia, diabetes, breathing problems, etc.) of a patient in a specific environment (e.g., smart homes, care centers, etc.). 
\end{enumerate}
Consequently, Things layer devices are not reliable and resilient to address key requirements of CPHS. For instance, they fail to reliably provide continuous monitoring of a patient's health which may threaten the patient's life. Furthermore, the devices also fail to support CPHS, as they are technologically personalized (i.e., for specific medical conditions (like blood glucose, blood pressure, heart rate, etc.), and are not clinically personalized (i.e., unaware for the other associated medical conditions of the patient). AI/ML-based approaches provides personalization in specific environment~\cite{HALEEM2021} and for specific diseases~\cite{zhang2020}.
\subsection{Communication Layer}\label{subsec:cl}
\begin{figure}[ht]
    \centering
    \includegraphics[scale=0.65]{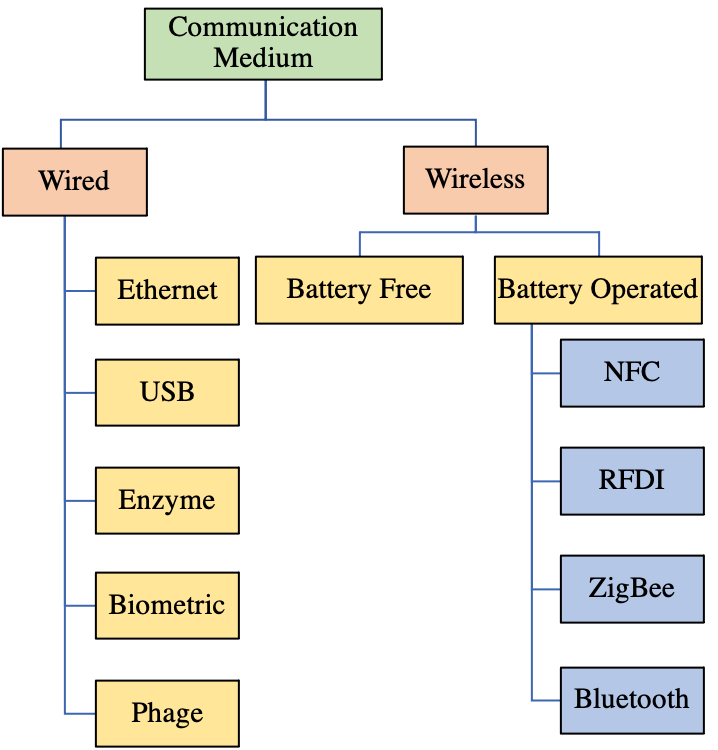}
    \caption{Classification of IoT Communication Mediums}
    \label{fig:cmedium}
\end{figure}

The communication layer is considered as the spine of the IoT systems~\cite{Mohamed2019}. IoT devices generate a large amount of continuous data that is challenging to store and communicate~\cite{aydos2019}. All the data from the Things layer is transferred to the application layer through the communication layer. A suitable communication medium is required to transfer the data from all the nodes of the Things layer to the other layers of the system. The communication medium could be wireless (WiFi, Bluetooth, RFID, 5G, etc.) or wired (Ethernet, USB, etc.) based on the designed communication protocol. Figure~\ref{fig:cmedium} shows the communication medium classification according to connectivity in an IoT system. 
% \begin{figure}[ht]
%     \centering
%     \includegraphics[scale=0.55]{figures/cmedium.png}
%     \caption{Communication Mediums in an IoT System}
%     \label{fig:cmedium}
% \end{figure}
 The major requirements for IoT communication in order to set up a successful healthcare 5.0 system include energy efficiency, range, cost, reliability, security, and scalability. 
%  \begin{itemize}
%      \item  Efficient from the energy point of view
%      \item Range
%      \item Cost effective
%      \item Reliable
%      \item Secure
%      \item No delay
%      \item Scalable
%  \end{itemize}
Real-time data collection and data processing are carried out at the communication layer on IoT platforms and in information systems~\cite{aydos2019}. 

As the communication layer in an IoT-based healthcare system is considered the backbone of IoT communication, it inherits all the problems of the communication network. Analyzing the communication layer problems, in the following, we
\begin{itemize}
    \item highlighted key challenges of the communication layer that hinders key requirements of HIoT as identified in the Sub-section~\ref{subsec:keyreqs},
    \item analyzed current approaches to address the above challenges, and
    \item identified the challenges as research gaps that have not been addressed adequately.
\end{itemize}
We discussed the above for each requirement, namely reliability, resilience, and personalization, in order.
\begin{enumerate}
 \item \textbf{Reliability:} 
According to the recent research~\cite{Moore2020} there are two forms of network reliability challenges,
which are:
\begin{itemize}
    \item Enhancement of QoS in the network
    \item  Evaluating the  reliability functions for networks
\end{itemize}
There have been various attempts for improving the reliability of IoT networks through enhancing QoS parameters and by quantifying components based on the monitoring. 
In~\cite{Almasri2018} the authors presented micro services QoS-aware middleware that monitors response time, throughput, availability, reliability, and cost parameters. In~\cite{li2017} authors used the Generalized Stochastic Petri Net (GSPN) approach for reliability modeling. The mathematical model used by them calculated the reliability of failure rate, consumption time, response time, and repair times. In~\cite{ sinche2018} the authors proposed a gateway redundancy model which reduced the RTT (return trip time) as the performance metric and proposed methodology lowers the return time to 1\%  from 14\% during fault conditions. In~\cite{Kamyod2018} the authors quantified some QoS metrics, like delay throughput, and packet loss. In~\cite{brogi2017} the authors proposed a general model for a QoS-aware IoT infrastructure, that only deals with latency and bandwidth. In~\cite{kumari2018} the authors provided an analysis of the role of fog computing, cloud computing, and the Internet of things for mitigating storage facilities related to the data of the patients and minimum capital investment on computing for uninterrupted context-aware services to the end-users. The authors proposed a three-layer patient-driven Healthcare architecture for real-time data collection, processing, and transmission.
In~\cite{tuset2020} the authors' combined packet replication with modulation diversity using multiple IEEE 802.15.4g SUN modulations for the evaluation of the benefits of improving communication reliability. Their results showed a significant increase in packet delivery ratio (PDR), that enhance the reliability of SUN networks for industrial applications. In~\cite{alam2020} the authors presented a framework for handling reliability in IoT based on the TCP (Transmission Control Protocol). The framework calculates delays, failure states, and re-transmission of the data.
In~\cite{akbar2021} the authors proposed a novel approach that uses using machine learning and multi-objective optimization (moo) for the software-defined network (SDN)-enabled adaptive and reliable communication in an IoT-fog environment. The authors successfully evaluated reliability using ML-based algorithms, while the "moo" algorithm is used to find the Pareto-optimal paths. The proposed approach is more faster and reliable compared to the other existing ones.

In~\cite{Humayun2020} the authors have provided a comprehensive framework to provide optimal solutions for privacy preservation, and energy optimization for researchers and practitioners in a better understanding of 5G infrastructures. The authors evaluated the proposed framework using case studies and mathematical modeling.

The afore-mentioned research has some limitations:
 \begin{itemize}
     \item The services are not scalable.
     \item Reliable performance only at a specific layer of the network like the edge layer.
     \item Limited view of network performance like limited to only TCP.
    %  \item Reliable only for the specific layer of IoT.
     \item Only specific for specific parameters of QoS (e.g., delay, throughput, packet loss, etc.).
 \end{itemize}
So far, there is no combined approach which deals with reliability considering both QoS and quantifying component parameters. 
 Being wireless, current communication infrastructures (e.g., media, protocols) used in HIoT are not reliable to support real-time and critical operations of CPHS. Due to interference, delays, and loss of data in wireless communication, data integrity becomes a serious problem from CPHS point of view mainly because a small variation in data can have significant consequences on a patient's health condition, e.g., a small error in blood pressure monitored values may not have a significant impact on the health condition, while a similar small error may have a significant impact on the health in case of monitored glucose values. Furthermore, to support continuous monitoring of health conditions requires high availability of the communication network, which is not the case in current wireless communication. However, 5G aims to support reliable communication~\cite{Ahad2019,Humayun2020} through ultra-low latency, high bandwidth, ultra-high reliability, reducing interference, and enhancing QoS. AI/ML-based approaches provide more faster and reliable communication in IoT-fog environments.
 %Another challenge from the reliability point of view is that in the case of self-configuration after software failure the system will lose the previous data and hence can lead to the wrong diagnosis. Some other issues related to the reliability of the communication layer are data integrity, availability of the channels at the desired time, etc., and many more.
 %%%%%%%%%%%%%%%%%%%%
 
 %%%%%%%%%%%%%%%%%%%%
    \item\textbf{Resilience:}
    Analogous to the Things layer, the resilience of the communication layer can be realized as the capability of communication systems to continue their normal operations and vigorous response in case of any unexpected or unpredicted situations. To support real-time resilience for IoT based healthcare systems is very challenging mainly because the resilience of such communication system arise from the complexity of failures, namely failures of 
    \begin{itemize}
        \item communication devices (e.g., router) or
        \item communication network (e.g., wireless media, protocol).
    \end{itemize}
    Former are partially resilient through reconfiguration, when current configurations get maligned but do not support resilience in case of any other mechanical failure. While the latter does not support resilience, for instance, any interference in wireless media and vulnerability in the communication protocol is not recoverable on the fly. Recently, various approaches have been proposed to support resilience in IoT-based communication networks. For instance, 
    in~\cite{Kim2019, Benson2018, Keef2018} authors have proposed different Cloud and Edge-computing-based mechanisms to support the recovery of various network failures in IoT networks including DoS attacks. However, they are limited in supporting only communication network failures due to denial-of-service (DoS) attacks and fail to support operational failures of communication devices in the face of other security attacks. Furthermore, there have been some efforts~\cite{Benson2016, Perez2016} that introduce resilient aware architecture and overlays to support device failures in the face of security threats. However, these approaches are retrospective, i.e., they are only resilient in known scenarios and fail to provide resilience in situations that are either variant of the known scenarios or different from the known.
   In~\cite{BABAR2020} the authors proposed secure demand-side management (DSM) engine that uses machine learning (ML) for the IoT-enabled grid that is responsible to preserve the efficient utilization of energy based on priorities. The proposed DSM engine reduces the utilization power of smart grids and is less vulnerable to intrusions. In~\cite{zeng2020}, the authors designed an authenticated key exchange (AKE) protocol that can resist side-channel attacks and enhance security and privacy in E-Health applications. The proposed protocol provides stronger security and higher efficiency compared to similar protocols in IEEE 802.15.6 and Bluetooth 5.0.
    
    Current communication infrastructures (devices and network) are typically not resilient, e.g., they fail to automatically recover in case of any soft (i.e., software/configuration failure/malfunctioning) or hard failure (i.e., hardware/device failure). AI/ML-based approaches can provide resilience by preserving the efficient utilization of energy based on priorities. They partially support resilience, for instance, soft resilience through self-configuration and hard resilience through configuring an alternative interface. However, these resilience methods hinder data integrity due to failure in recovering the loss of pre-configuration/failure data.
    %\todo{stateful resilience is a key requirement of HIoT - a system is capable of recovering in a state that establishes data integrity even for the data loss during the recovery time.}
    \item \textbf{Personalization:} The Communication layer supports personalization by providing various personalized communication networks (e.g., Wireless Personal Area Network - WPAN, Wireless Body Area Network - WBAN, and Internet of Medical Things - IoMT) that connect various medical devices of an individual patient. Although these networks support data and communication protection among personal devices, it fails to protect data and communication when these networks interact with external networks (i.e., Internet) for monitoring and analysis of patients' health conditions. For instance, in~\cite{Gordana2020, Ghub2020, Hari2020}, various personalized communication mechanisms have been proposed that enable secure transmission of information among various medical devices. However, these approaches are limited in enabling on the fly integration and communication of new devices (e.g., IoT medical devices from different manufacturers supporting different communication protocols), and in protecting the end-to-end transmission of data among external healthcare agents (e.g., practitioners, carers, and nurses) that involve other networks (e.g., Internet, LAN, and WAN).
    
    Contemporary IoT-based communication infrastructure achieves personalization through supporting customized communication configuration for required healthcare services. For instance, 5G aims to support the traffic-free transmission channels for real-time data exchange among HIoT components to support emergency medical services. Furthermore, personalization can also be supported using personal clouds to prioritize the desired services.
\end{enumerate}
%\textcolor{red}{Reliability, resilience and personalization benchmarks or metrics, e.g., ISO, HL7}

Consequently, the communication layer is not reliable and resilient to address key requirements of CPHS, for instance, the current communication infrastructure fails to reliably exchange data for continuous monitoring of patient's health which may threaten the patient's life. Furthermore, the communication also fails to support CPHS, as they are personalized in a way that supports traffic-free transmission channels for dedicated services that are not stateful, i.e., the channels are not aware of the significance/semantics of the transmitted data generated through monitoring different medical conditions of the patients. 
    %\todo{current communication channels do not support prioritization of communication based on the sensitivity of the data}
\subsection{Application Layer}\label{subsec:al}

\begin{figure*}[ht]
    \centering
    \includegraphics[scale=0.75]{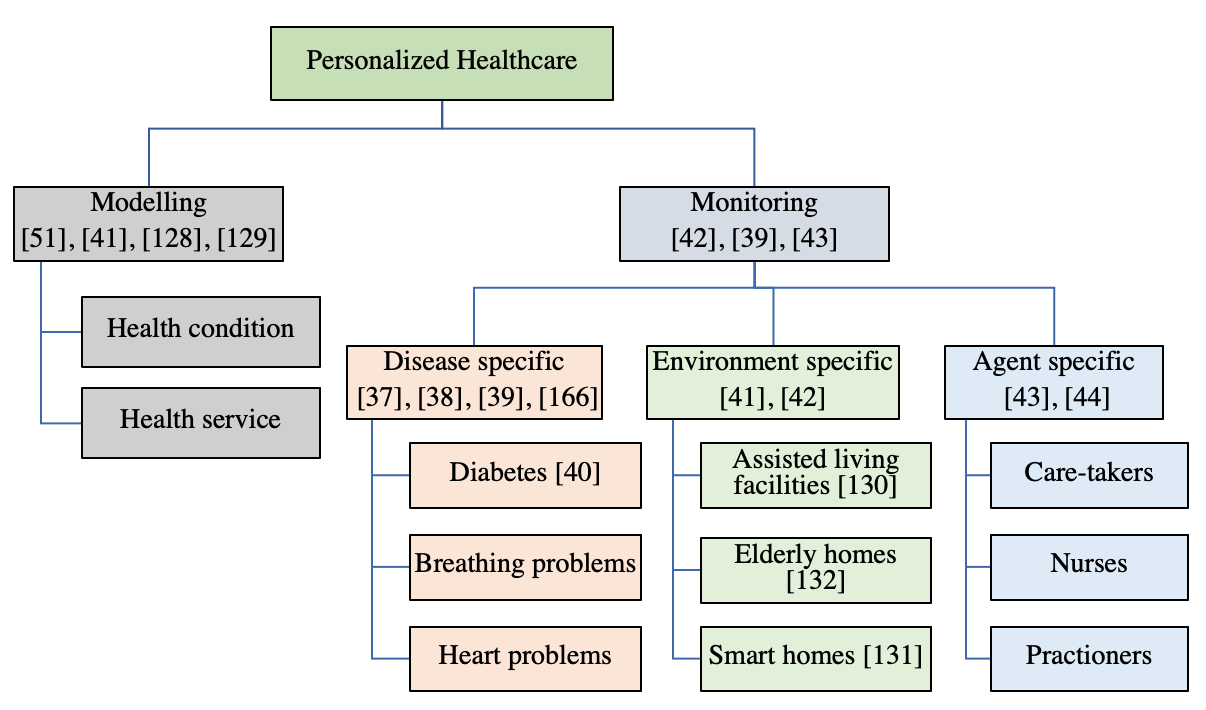}
    \caption{Classification of Personalized Healthcare Modeling and Monitoring Techniques}
    \label{fig:phmc}
\end{figure*}

The application layer communicates the users
with the IoT/Internet of Medical Things (IoMT) platform directly and focuses on providing high-quality services in healthcare. IoMT is a connected infrastructure of devices, software, hardware, and services. With the advancements in connectivity, their use is exponentially increasing. These are used by healthcare professionals for processing and analyzing data for decision-making and patient treatment.
 In hospitals, the IoMT is typically about increasing patient safety and/or optimizing processes. It enables practitioners to work together across disciplinary boundaries to carry out individualized patient care. The goal of the application layer is two-fold, it provides:
\begin{itemize}
    \item \textbf{Monitoring Applications:} that process the data and supports decision making on one hand, and
    \item\textbf{Support Applications:} that are mare interfaces to operate various health monitoring devices on the other hand.
\end{itemize}
 Different approaches have been developed to support personalized healthcare services. Most of the approaches support personalized healthcare services by using various healthcare modeling and monitoring techniques as shown in Figure~\ref{fig:phmc}. 

\subsubsection{\textbf{Healthcare Modeling}}
There have been various efforts where personalization is addressed in the frame of healthcare modeling and monitoring. For instance in~\cite{Wang2018} the authors studied and investigated the modeling method for a cloud healthcare system. In this system, the authors used the telemedicine platform to share high-quality medical resources among big tertiary hospitals and small community hospitals. Their presented Petri net model used the colored Petri nets (CPN) tools for verifying and analyzing the simulation process. The model describes the state of patients and the relationship between the medical process and resources in this cloud healthcare system. 

In~\cite{Abdulrauf2018}, the authors focus on the formal model Petri net where public-key encryption controls the security model. The authors successfully used Petri net mathematical modeling for solving graphical notations and minimizing the risks and overcoming the security risks in the healthcare industry.
% In \cite{Minskey2010} the authors addressed the dangers of privacy and integrity of the electronic records during electronic transfers by formulating and enforcement of the social policies and laws about the exchange of electronic healthcare records between large and heterogeneous networks of healthcare providers. The authors introduced a decentralized, seamless, and secure inter-operation between different healthcare providers under different policies and domains and provide support for the naturally hierarchical organization of these policies that govern the exchange of healthcare records. 
In \cite{Yafei2018} the authors focus on cloud-based healthcare systems. The authors presented Petri net for describing the relationship among medical processes and resources in their integrated healthcare system. The model solves the patient assignment scheduling problem using a greedy-based heuristic algorithm. 
In~\cite{jovanovic2019} the author employed machine-learning techniques to extract features from electrocardiogram (ECG) and informed hypertensive patients about their heart status during various activities. The experimental data were collected from a network of conscientious volunteers through a mobile crowdsensing (MCS) application. However, the experimental data did not include the main diagnostic feature for hypertension that is blood pressure. In~\cite{FOUAD2020} the authors used an IoT sensor with artificial intelligence (AI) to predict the exact patient details such as fitness tracker, medical reports, health activity, body mass, temperature, and other health conditions. The collected patient details are processed according to the iterative golden section optimized deep belief neural network (IGDBN) which helps to choose the effective patient assistance system.
In~\cite{hathaliya2019} the authors integrated Machine learning (ML) and blockchain technologies to solve the basic challenges of securely exchanging the big data that supports the analytics. The authors used ML algorithms like Naive Bayesian, k-nearest neighbor, and decision tree on the trained data to achieve higher accuracy in identifying the disease.
In~\cite{ amin2020edge} the authors analyzed health data classification by tracking and identifying vital signs using state-of-the-art deep learning techniques and highlights the challenges (limited computational power, data redundancy) posed by edge intelligence.
In~\cite{ vaananen2020survey } the authors surveyed the practicality of AI services for healthcare by healthcare professionals. The authors provide an analysis of the healthcare services that are using AI technology. According to the survey results majority of the health professionals are willing to use AI technology for supporting decisions.

 \subsubsection{\textbf{Healthcare Monitoring}}

In~\cite{Corno2016} authors explored and applied the IoT paradigm in the context of assisted living facilities (ALFs). They have designed and implemented a system capable of supporting the healthcare assistants in their daily livings while operating in ALFs with physical or cognitive disabilities. The system used wearables and mobile technologies to attain its goal. The system supports both inhabitants with cognitive disabilities with instant assistance through their wearable devices when required. The caregivers in potentially hazardous situations are alerted automatically. However, their focus is on one specific group with physical and cognitive disabilities.
In~\cite{mano2016}, the authors exploited IoT technologies for enhancing healthcare in smart homes through patient identification and emotion recognition. They have automatically used images and emotions by using IoT infrastructure for helping healthcare in the smart home environment. Images were used for the detection of the right person to ensure to provide the right treatment while emotions are used as they play a very crucial role in the recovery process. They have used facial angles, gestures, and postures for emotion recognition. Smart watches and clothes with wearable sensors can be included in the smart home environment as a means of collecting physiological information which is then used for decision making.

In~\cite{ kaur2017ai} the authors used AI-based predictive and prescriptive analysis by using open source technologies like Apache beam, Apache Flink Apache Spark, Apache NiFi, Kafka, Tachyon, Gluster FS, NoSQL- Elasticsearch, and Cassandra for data processing and used it for extracting useful knowledge that helps in decision making and patient’s monitoring.
In~\cite{AJara2013} the authors presented an architecture for continuous monitoring, ubiquitous connectivity, extended device integration reliability, security, and privacy support. The continuous monitoring is made feasible with the use of a proposed protocol called YOAPY. The efficiency and security are scalable integration of sensors that are deployed in the patient's personal environment. The focused group of patients is the patients with breathing problems.
As the HIoT are continuously providing information about the health conditions of the patients, a large amount of data is collected relating to the patient's health conditions which enable the use of advanced machine learning techniques. These techniques can be used to identify the information about user data, the effect of the treatment on patient's health, and monitor the progress~\cite{Garda:2020}. In~\cite{esteva2017} the authors used machine learning (an AI technique) for the classification of skin lesions using a single convolutional neural network (CNN), trained end-to-end from images directly, using only pixels and disease labels as inputs. The classification was able to successfully diagnose skin cancer as efficiently as dermatologists.  
In~\cite{Ali:2020} the authors proposed a smart healthcare system for heart disease prediction using ensemble deep learning and feature fusion approaches. The feature fusion method generates valuable healthcare data (excluding irrelevant and redundant features) from sensors and medical healthcare records.
 According to~\cite{Sloane:2020} AI-based systems can significantly reduce delays while incorporating new best practices and tools, for new significant periodic revision of scientific and medical knowledge in healthcare systems and help to optimize output more quickly.

In~\cite{Ahmed2015} authors proposed a personalized self-served health-monitoring system for the elderly in the home environment by combining general rules with case-based reasoning (CBR) approach. The system uses health parameters such as blood pressure, blood glucose, weight, activity, pulse, etc., and applies general rules to classify individual parameters. Case-based reasoning is used to combine different health parameters which ultimately generates the overall classification of health conditions. The system successfully generates feedback, recommendations, and alarms in a personalized manner. However, the system works only in the specific elderly home environment and for a specific age group. Moreover, the system uses rule-based reasoning (RBR) and case-based reasoning (CBR) for the classification of health conditions. Facts and rules are strictly categorized in RBR and require a strict Boolean match on the conclusions and suppositions while the real-world problems like healthcare problems are often fuzzy and do not match exactly with rule-based conclusions and suppositions while case-based reasoning has its limitations w.r.t, the representational scheme used for the cases in the case library, structure of case library,need-based,retrieval and similarity metrics, and adaptation.

In~\cite{ALI2016} the authors propose a multi-modal hybrid reasoning methodology (HRM) that integrates the rule-based reasoning, case-based reasoning, and preference-based reasoning approaches sequentially by exploiting guideline rules, past successful experience cases and personal preferences to generate personalized physical activity recommendations according to the user's needs and preferences. The authors had successfully facilitated the users with different wellness services. However, the used approaches are not exhaustive and are dealing only the wellness services at a personalized level but not the personalized health of the users. 
 
In~\cite{Alfian2018} the authors proposed a personalized healthcare monitoring system for diabetic patients by utilizing Bluetooth Low Energy (BLE) based sensors and real-time data processing. Real-time data processing utilizes Apache Kafka for handling incoming sensor data whereas MongoDB is utilized to store the unstructured sensor data. The authors successfully handled continuous data (e.g., Blood Glucose (BG), heart rate, blood pressure, weight, and other personal data) from the BLE-based sensor devices by real-time data processing. The authors used multi-layer perceptron (MLP) to classify diabetes patients.
In~\cite{ma2017} authors proposed a smart application that monitors the health status by an easy user interaction for the users with limited technological experience. To implement the system, authors combined the Wearable sensors such as the Angel Sensor (for data collection) with voice interactive devices, and the Amazon Echo (for voice interfacing).
In~\cite{jovanovic2019} the author employed machine-learning techniques to extract features from electrocardiogram (ECG) and informed hypertensive patients about their heart status during various activities. The experimental data were collected from a network of conscientious volunteers through a mobile crowdsensing (MCS) application. However, the experimental data did not include the main diagnostic feature for hypertension that is blood pressure. 
In~\cite{FOUAD2020} the authors used an IoT sensor with artificial intelligence (AI) to predict the exact patient details such as fitness tracker, medical reports, health activity, body mass, temperature, and other health conditions. The collected patient details are processed according to the iterative golden section optimized deep belief neural network (IGDBN) which helps to choose the effective patient assistance system.
In~\cite{esteva2017} the authors used machine learning (an AI technique) for the classification of skin lesions using a single convolutional neural network (CNN), trained end-to-end from images directly, using only pixels and disease labels as inputs. The classification was able to successfully diagnose skin cancer as efficiently as dermatologists.

%%%%%%%%%%%%%%%
The afore-mentioned approaches lack in one or many of the followings:
\begin{itemize}
    \item They establish personalized healthcare service through visualizing personalized data that is focused on monitoring a selective health condition (like for patients with breathing problems, diabetes mellitus). The systems provide security and privacy support but fail to monitor the comprehensive personalized health condition of the patient.
    \item They applied IoT environment in a specific context (like assisted living facilities, smart homes, cognitive disabilities). The system only assists the caregivers and inhabitants in hazardous situations but is unable to monitor the comprehensive health condition of inhabitants. 
    % \item They proposed a system that is working in a specific environment e.g., smart homes for specific diseases where emotions and images are playing their role thus failing in monitoring the comprehensive personalized health of the patients.
    \item They developed a system that is working for people with less technological experience by using voice interactive devices in combination with smart sensors like Angel sensors. The system works for specific health conditions like diabetes mellitus and is not able to monitor the comprehensive personalized health condition of the patient. 
    % \item They present models with patients' personalized environment (like for patients with breathing problems). The systems provide security and privacy support but are not able to monitor the comprehensive personalized health of the patients.
    % \item They design and implement systems for the caregivers who monitor the specific group of people with physical and cognitive disabilities.
\end{itemize} 

In the following, we have highlighted key challenges of the application layer that hinders key requirements of HIoT as identified in the Sub-section~\ref{subsec:keyreqs}, analyzed current approaches to address the above challenges, and identified the challenges as research gaps that have not been addressed adequately. We discussed these challenges for each requirement, namely reliability, resilience, and personalization, in order below:
\begin{enumerate}
 \item \textbf{Reliability:} The main goal of healthcare applications is to derive various facts about a patient's health based on monitoring through collecting data about a patient's health conditions from respective medical devices. The derivations are later used by the practitioner, patient, or caretakers to control the patient's health. Therefore, the reliability of these applications strictly depends on the reliability of such derivations.
 
 Analogous to the Things and communication layer, the reliability of the application layer can be defined considering the following three requirements:
 \begin{itemize}
     \item Perform a required function.
     \item Perform the function under stated conditions.
     \item Operation for a specific time. 
 \end{itemize}
 In the following different approaches have been established/ developed which address one or more reliability requirements of the applications layer.
 In~\cite{vergutz2020} the authors developed the reliability of the healthcare system, investigated and quantified their influence of the component states (functioning levels and failure) based on unified methodology. 
In~\cite{marcelis2020} the authors investigated the long-range wide-area network (LoRaWAN) frames loss due to the channel effects. They designed a novel coding scheme that combines techniques from convolutional and fountain codes called DaRe that works on the application layer for data recovery and provides better resilience.
%to bursty frame losses.

According to~\cite{peter2006} reliably helps to deliver effective interventions in healthcare organizations.They proposed a model to improve reliability which includes interventions to improve by focusing on valid rate-based measures. They used the following interventions for improving the reliability of the healthcare system:
\begin{itemize}
    \item Identifying evidence-based interventions,
    \item selecting interventions with the most impact,
    \item developing measures to evaluate reliability,
    \item measuring baseline performance, and
    \item ensuring that patients receive the evidence-based interventions.
\end{itemize}
In~\cite{timothy2016} the author examined an innovative approach that uses the combination of specific work practices and behavioral processes for detecting unexpected events and then operates in a nearly error-free manner. Furthermore, they explored that reliability-enhancing work practices (RewPs) help to improve the performance of the system by reducing errors. Also, the RewPs are directly and indirectly associated with fewer medication errors and patient falls.
 In~\cite{vyas2019} the authors used the features of decentralization of database and consensus in Blockchain technology to get highly accurate results in terms of machine learning with the security and reliability of Blockchain Technology.

 In~\cite{vergutz2020} the authors presented fingerprinting healthcare applications architecture based on network slicing that can provide reliability for s-health applications and services. The developed system obtains 90 percent accuracy assisting in network customization. 
In the article~\cite{John2019} authors include purpose, process, people, and management system as the important constructs which are required for highly reliable healthcare. The article also explores the reasons for the failures in improvements in healthcare and suggests rigorous training and continuous attention for updating the standards. 
In~\cite{hathaliya2019} the authors integrated Machine learning (ML) and blockchain technologies for solving the challenges of securely exchanging the big data that supports the analytics. The authors used some ML algorithms like Naive Bayesian, k-nearest neighbor, and decision tree on the trained data to achieve higher accuracy in identifying the disease.
In~\cite{shi2020} the authors emphasized the great significance of the convergence of IoT and AI. they elaborated the knowledge-enabled AI and data-driven AI, comparing its advantages and disadvantages throughout the IoT architecture, from the things layer through the communication layer to the application layer. The authors highlighted that reliability in data transmission, security and privacy are open issues during the convergence between IoT and AI.
In~\cite{singh2021} the authors presented blockchain-based electronic healthcare record (EHR) using javascript-based smart contracts for the patient-centric design of a decentralized healthcare management system. The authors developed and implemented a prototype based on hyper ledger fabric that provided performance such as latency, throughput, resource utilization for varied scenarios, and control parameters.

\item\textbf{Resilience:} Analogous to things and the communication layer, the resilience of the application layer can be realized as the capability of applications (e.g., healthcare service monitors and controllers) to continue their normal operations and vigorous response in case of any unexpected or unpredicted situations. To support real-time resilience of IoT based healthcare applications is very challenging mainly because applications may fail due to:
    \begin{itemize}
        \item Inadequate and vulnerable design of applications (e.g., techniques that application employ (for instance, machine learning and AI), programming languages used to develop applications (for instance, Python and Java), to name a few), and
        \item inadequate and vulnerable applications' execution (e.g., data, network, and run-time libraries).
    \end{itemize}
Former (aka design) failures are key threats to the resilience of healthcare applications being unrecoverable in real-time due to the time required to localize the failure and then fix it. Furthermore, design limitations (e.g., false alarms, bias, complex processing) of techniques (e.g., machine learning and AI) that are used in healthcare applications may lead to severe consequences resulting in life threats to patients~\cite{David2019}. Different approaches~\cite{Maity2017, Tobore2019, Kelly2019, Jiang2017} have sketched various limitations of application techniques in healthcare and have devised ways to address them in a way that requires developers to handle various resilience concerns while developing such applications, e.g., dataset shift, accidental fitting of confounders, unintended discriminatory bias, generalization to new populations, and the unintended negative consequences of new algorithms on the health condition outcomes. However, these approaches are error-prone being manual, and do not ensure application resilience. More recently, model-driven approaches have been used to significantly improve the design of such applications to support resilience. For instance, in~\cite{Mez2017, Shamout2021, Yu2019, Ravi2017} various models have been proposed to design such applications that do not include different vulnerabilities and inadequacies supporting resilience by design. However, these approaches only support resilience in the known (via models) scenarios, and fail to recover from failures that are either variants of the modeled ones or different from them. 

Latter (aka execution) failures are harder to recover mainly because they involve unseen incidents that take a lot of time to identify the incidents' cause and then to fix them. Furthermore, identification here is challenging because the execution environment involves very complex infrastructure most of which is a black box (e.g., communication infrastructure and IoT devices from different manufacturers and platforms and services from different providers). To this end, various monitoring techniques have been developed to detect such incidents at run-time and mitigation techniques that recover the application's failures in a way to minimize the impact of the incident. Current approaches~\cite{Rod2021,Camara2016,Vala2018} introduce different risk-based maintenance methodologies that enable recovery of healthcare applications making them resilient aware in certain scenarios against known threats.
In~\cite{Venkataramani2020} the authors presented a novel and resilient approach for improving the operational efficiency of computational cost by using approximate computing (AxC) that can significantly boost the efficiency for adopting AI-based applications and services. They presented a multi-tera operation per second (TOPS) AI hardware accelerator core that they built from the ground-up using approximate computing (AxC) techniques across the stack including algorithms, architecture, programmability, and hardware.

\item\textbf{Personalization:} Contemporary healthcare applications support personalization~\cite{Choe2019, River2019} in one or more of the following different ways:
\begin{itemize}
    \item Patient's self-organization of various health conditions,
    \item health condition and/or patient-specific gadgets and automated guidance. 
    % \item health condition and/or patient-specific automated guidance.
\end{itemize}
The patient's self-organization is supported by developing applications~\cite{Tao2015, Gai2019} that can be customized as per patient's requirements, e.g., goal setting, celebration, discovery, reflection, and coordination among others. However, these approaches do not support clinical personalization that requires clinical optimization of various health conditions to reduce health risk and to improve the quality of health/life of the patients.

Various gadgets (and associated applications)~\cite{Kim2020, Sheth2017, Hong2020, Lin2020} have been developed to support the customized organization of various health conditions. These gadgets provide health condition-specific monitoring and control patiently. These approaches provide very restricted customization of gadgets for a specific health condition of the patient.

Recently, various ML and AI-based applications~\cite{Knick2018, Ketan2020, Rowe2020, Emilio2020, Ahmed2020} have been developed that provide personalization of a specific health condition to the patient based on data collected from personal medical devices of the patient over long periods. However, the above-mentioned approaches are limited in realizing the personalization that is data-driven which suffers from false predictions resulting in a severe loss.
In~\cite{Paranjape2020} the authors addressed the major issues such as explainability, liability, and privacy to mainstream AI in healthcare. Explainable AI, a new emerging discipline seems to be the solution for making machine decisions transparent, interpretable, traceable, and reproducible.

\end{enumerate}
Consequently, the application layer supports either technological personalization or health condition-specific personalization to patients. This support is limited and does not provide clinical personalization of various health conditions of a patient in a way that reduces the health risk on one hand and improves the quality of life and health on the other hand. AI/ML-based approaches can be partially used at the application layer. These techniques can be used for health condition monitoring or for dealing with the patient data e.g., patient's history, treatment record, security of patient data, and progress about recovery. But for critical matters like diagnosis, treatment AI/ML-based techniques can not be implemented due to the explainability and trust of these applications. A lot of work is still in progress regarding the explainability and trustworthiness of AI. 

%%%%%%%%%%%%%%%%%%%
\begin{table*}[ht]
\caption{Research Gaps for CPHS Requirements}
\centering
%\begin{tabular}{|l | l |l| l|}
\begin{tabular}{|>{\centering\arraybackslash}m{2cm}|>{\centering\arraybackslash}m{4cm}|>{\centering\arraybackslash}m{4cm}|>{\centering\arraybackslash}m{4cm}|}
\hline
\hline
\rowcolor{gray!20}
\diaghead{\theadfont Diag Columnmn Head II}%
  {\bf{IoT Layers}}{\bf{Requirements}} & \textbf{Reliability} & \textbf{Resilience} & \textbf{Personalization}\\
 \hline \hline
\textbf{Things } &  \specialcelll{Discrepancies by design\\ Maligned software\\ Mechanical failure\\Limited resources} & \specialcelll{Fail to recover\\ Limited hard resilience}& \specialcelll{Limited personalised\\ healthcare services\\ disease specific }\\
\hline

\textbf{Communication } &  \specialcelll{No reliable real-time support\\No reliable support for \\ critical operations\\ No reliable data exchange} & \specialcelll{No automatic recovery in \\case of failure} & \specialcelll{Personalised support only \\in case of personal clouds}\\
\hline
\textbf{Application } &  \specialcelll{False detection due to ML\\ Not rigorous} & \specialcelll{Inadequate  and  vulnerable\\  design  of  applications\\ Inadequate  and  vulnerable\\  applications} & \specialcelll{Support technological\\ personalization\\ Health condition specific\\ personalisation }\\
\hline
 
\end{tabular}
\label{table:gaps}
\end{table*}
%%%%%%%%%%%%%%%%%%%
\section{Research Gaps}
The research gaps in the IoT layers in the frame of comprehensive personalized healthcare services requirements are shown in Table~\ref{table:gaps}. These derived gaps determine research challenges that are associated with different layers of the architecture and corresponding key requirements of healthcare. The identified gaps conclude that the current HIoT devices and applications monitor a specific health condition of patients\cite{IoTmarket2020} (e.g., insulin level, ECG level, and oxygen saturation monitoring devices) and thus are not reliable for a patient with multiple health conditions as they fail to provide reliable, resilient, personalized, and application-specific healthcare mainly because these applications,
\begin{itemize}
    \item do not coordinate with other similar applications that monitor different health conditions of the same patient and as a result, fail to understand the health context of the patient, and,
    \item do not exchange information securely due to unreliable and heterogeneous underlying software and hardware.
    \item does not understand the effect and relation among different health conditions of a person. 
\end{itemize}
% \subsection{\textbf{Research Questions}}
% \begin{itemize}
%     \item How to ensure that the coordination (i.e.,
%     information collection and sharing) among healthcare IoT applications monitoring different health conditions of the same patient is reliable?
%     \item How to identify the right cause of a certain health condition while taking into account the interdependencies of multiple health conditions and coordination of different health condition monitoring IoT devices and applications?
%     \item How to identify the right treatment for the identified cause of the health condition minimizing (if not eliminating) risk of the treatment in general and its side-effects on other health conditions?
% \end{itemize}
The comparison of various smart healthcare approaches like AI-based, Machine learning-based, and deep learning-based is shown in Table~\ref{table:comparisons}. The comparison clearly shows that only few among the above mentioned approaches  either partially support (i.e., indicated by half filled circle) or do not support (i.e., indicated by empty circle) personalization of healthcare services considering multiple health conditions of a patient. 
The above mentioned analysis of the CPHS requirements (i.e., reliability, resilience, and personalization) at different layers has been performed under the assumption that operations of the respective layers exhibit normal behavior, which is not true in practice due to emerging security threats to different layers. Therefore, we consider security as integral to the CPHS requirements that enable us to identify the security threats that affect the normal behavior of the layers. In the following section, we discuss various security threats to different IoT layers.

%%%%%%%%%%%%%%%%%%%%%%%%%%%%%%%%%%%%%%%%%%%%%%%%
\section{Security Threats at IoT Layers}\label{sec:s_threats}
Security threats at different IoT layers and their possible mitigation are shown in Table~\ref{tab:IoTsecurity}.
\subsection{Security of IoT Devices}\label{subsec:tlt}
The increase in the number of modern IoT devices and their exponential use in healthcare systems has increased the risk of security-related issues and other vulnerabilities. As per \cite{Cisco20} till 2025 number of IoT devices per second will rise to 152K. The emergence of Wi-Fi 6 and 5G networks provides higher data rates to support new applications and increases network capacity. They are exponentially scalable and will bring much better performance for organizations with the improvement in speed, latency, and scalability and will be ideal for indoor enterprise networks. The main reasons associated with security issues are as follows:
\begin{itemize}
    \item The nature and sensitivity of the data which is communicated by medical devices.
    \item Complexity and data compatibility issues.
    \item Lack of attention towards security issues by manufacturers of medical devices.
\end{itemize}
Due to the above-mentioned reasons, security-related issues related to confidentiality, integrity, and availability (CIA) are increasing~\cite{Somayeh2019} for IoT devices.\\
These devices are not fully secure and face different types of physical attacks. Some examples of these attacks are as under ~\cite{yang2017}~\cite{Tuhin2015}:
    \begin{itemize}
            \item Node tampering.
            \item RF interference on RFID's.
            \item Node jamming in Wireless Sensor Networks. 
            \item Malicious node injection.
            \item Physical damage.
            \item Social engineering.
            \item Sleep deprivation attack.
            \item Malicious code injection.
       \end{itemize}
According to~\cite{mahmood2015}, following security issues exist in the IoT Things layer:
\begin{itemize}
    \item strength of wireless signals,
    \item node interception by owner/attackers, and
    \item nature of network topology.
\end{itemize}
In addition to this, Replay Attack can easily exploit the confidentiality of this layer by spoofing, altering, or by Timing Attack. Another attack is the Node capture attack which can be made by the attacker by taking over the node and capturing all the information on the node. According to~\cite{farooq2015} another attack called DoS attack is also possible where the attacker adds another node in the network and threatens the integration of the data. This attack deprives the system of the sleep mode that the nodes normally use to save energy. Cellular networks have started to support new user categories tailored for IoT applications, e.g., narrow band-IoT (NB-IoT). Several challenges arise with the increase in the density of IoT devices. Re-transmission gets more frequent as collision among the IoT devices also increases due to several access requests for IoT devices which affects the energy efficiency. Several approaches are used to aggregate IoT traffic. For instance,
in~\cite{hattab2020} the authors proposed a transmission protocol to aggregate IoT traffic by using drones along with the surety of fair shared spectrum access with the existing cellular users.

\subsection{Security of IoT-based Communication}\label{subsec:clt}
The communication of IoT devices is the target for harmful attacks in the following three classical aspects~\cite{furdek2016}.
 \begin{itemize}
     \item\textbf{Confidentiality:} specifies to the communication to be accessed only by authorized persons.
     \item \textbf{Integrity:} refers to the complete trustworthiness transmission and receivable of data without any accidental and intentional modifications. 
     \item \textbf{Availability:} is about making sure about the usability of the service upon the authorized request only.
 \end{itemize}
 
The communication layer is not fully secure and can suffer from the following network attacks (attacks over the network or an attack on the network)~\cite{yang2017}~\cite{Tuhin2015}. 
 \begin{itemize}
 
 \item\textbf{Man in the Middle (MITM):} In this attack, the attacker capture, read, and modify data between two objects on the IoT platform ~\cite{farooq2015}. The major goal of the attacker is to change data on the IoT platform by sabotaging the traffic.
 \item\textbf{Spoofing:} In spoofing the attackers imitate or modify the node information to poison the network traffic by generating new nodes, extending or shortening the network paths, or by sending false messages~\cite{nawir2016,farooqumer2015}. Spoofing affects the integrity of the IoT system.
 \item\textbf{Desynchronization:} In desynchronization, the attackers interfere with the communication parameters and disturb the normal synchronous network traffic~\cite{virat2018, rostampour2018}. 
 \item\textbf{Selective Forwarding:} In selective forwarding, the attacker steals and corrupts the data by seizing the nodes, reducing the number of data packets, and redirecting the packets to different directions in IoT networks.
 \item\textbf{Unfairness:} In unfairness, the attackers distract the equal load sharing of the network, and as a result system gets down.
 \item\textbf{Flooding:} In flooding, the attacker disables a part of the whole IoT network and results in reducing the speed or the complete shut down of the service.
\end{itemize}
 Moreover, the other attacks include wormhole, Sybil, RFID cloning, unauthorized access, sinkhole attacks, denial of service, and routing information attacks, etc.
 %%%%%%%%%%%%%%%%%%%%
\begin{table*}[ht]
\caption{Security Threats at Different IoT layers~\cite{singh2020}}
\centering
\begin{tabular}{|>{\centering\arraybackslash}m{2cm}|>{\centering\arraybackslash}m{6cm}|>{\centering\arraybackslash}m{6cm}|}
\hline
\hline
\rowcolor{gray!20}
 \textbf{IoT Layers}  &  \textbf{Attacks} & \textbf{Potential Mitigation}   \\ 
 \hline
 \hline
\textbf{Things} & \specialcelll{Tampering, Jamming\\ Eavesdropping\\ Denial of Services\\Physical Attacks\\ Exhaustion} & \specialcelll{
Tamper-resistant packaging\\Channel surfing , priority messages\\End-to-end Encryption\\ Spread-spectrum techniques\\Shared cryptography\\Keys or routing tables\\Rate limitation\\ }\\ 
\hline
\textbf{Communication }&  \specialcelll{MITM, Spoofing\\ Desynchronization\\ Selective forwarding\\ Unfairness, Wormhole, Sybil\\ Flooding} & \specialcelll{Security monitoring\\token-based authorization\\ Semantic network firewall\\Traffic control, Link Authentication\\Client puzzles}\\
\hline
\textbf{Application} & \specialcelll{Phishing attacks\\ Virus, worms, Trojan horse\\ spyware, Malicious scripts\\Denial of Services\\ Side channel attacks\\ Crypt-analysis, MITM Trojan } & \specialcelll{Security training\\antivirus\\Traceability\\Validation } \\		
\hline
\end{tabular}
\label{tab:IoTsecurity}
\end{table*}
 %%%%%%%%%%%%%%%%%%%%
 \subsection{Security of IoT Applications}\label{subsec:alt}
 On the application layer users and machines communicate directly to the IoT platforms. Application layers are not fully secure and face different types of attacks are listed under.
 \begin{itemize}
     \item\textbf{Software Attacks:}  IoT applications face security vulnerabilities like,
     \begin{itemize}
        \item Phishing attacks,
        \item Virus, worms,
        \item Trojan horse,
        \item spyware,
        \item Malicious scripts and
        \item Denial of Service
     \end{itemize}
     \item\textbf{Encryption Attacks:} Along with the software attacks there are encryption attacks that break the system encryption. Some examples are as under:
    \begin{itemize}
        \item Side channel attacks
        \item Crypt-analysis attacks
        \item Man in the Middle attack
     \end{itemize}
 \end{itemize}

\subsection{AI-based Security Approaches}\label{subsec:ai}
In~\cite{Heka2020} the authors demonstrated vulnerabilities to different cyberattacks on personal medical device communication. Specifically, how an external attacker can hook into the personal medical device's communication and eavesdrop on the sensitive health data traffic, and implement man-in-the-middle, replay, false data injection, and denial-of-service attacks. The authors proposed an intrusion detection system (IDS), HEKA, for monitoring device traffic and attacks. The proposed system detects irregular traffic-flow patterns using an n-gram based approach and different machine learning techniques for intrusion detection.
% 1. AI as a threat (e.g., un-reliable, black box, unfair, biased, privacy)
 Undoubtedly, Artificial intelligence (AI) is revolutionizing patient healthcare with its seemingly limitless power. However, the new issues arising from the applications of AI in healthcare can not be ignored. In~\cite{price2019} the authors highlighted the new issues of AI-based applications in healthcare. Big patient data brings big risks about data liability and challenges about patient privacy concerns.
In~\cite{martinez2018,lamanna2018} the authors highlighted concerns about cybersecurity, the question of responsibility, and consideration of ethics while integrating AI tools into current practice. In authors~\cite{lamanna2018} addresses four major areas like transparency, informed consent, privacy, and accountability, where guidelines and best practices will be helpful to avoid harmful effects or unwanted consequences.
In~\cite{dasoriya2018} the authors discussed the uncertain future of AI considering its advantages and disadvantages. AI-based techniques can be used in different fields to achieve excellent results. For instance, according to \cite{kumar2016} AI-based robots and machines are replacing laborious and monotonous tasks. AI-based health reports have made doctors work faster and help in providing earlier diagnoses by reducing costs.
%%%%%%%%%%%%%%%%%
% 2. AI as defense (e.g., security monitoring, intrusion detection)
In~\cite{hasan2019} the authors highlighted current security solutions with AI i.e., Intrusion Detection and Prevention Systems (IDPS) as well as their limitations and considerations in a power utility network. 
In~\cite{yasmin2019} the authors presented artificial intelligence analysis benefits in medicine. The authors examined the effect of artificial intelligence assistance in the medical field, the effect of AI-based diagnosis on a patient, patient's treatment (i.e., Precision Medicine), error reduction (i.e., human and machine-based errors), and virtually being present with the patients(i.e., robot nurses, medicine reminders, etc.).
In~\cite{Xiao2018} the authors investigated the attack model for IoT systems and IoT security solutions based on machine learning (ML) techniques including supervised learning, unsupervised learning, and reinforcement learning (RL). The main focus was IoT authentication, access control, secure offloading, and malware detection schemes to protect data privacy.
In~\cite{ esmaeilzadeh2020use} the authors' surveys about the use of AI-based tools for healthcare purposes from consumers’ perspectives. According to the author acceptance of AI technology in the healthcare sector requires a deep understanding of the technology and related factors (ethical, regulatory concerns) before its implementation in clinical decision support (CDS).

Based on the identified gaps and the security threats on the three layers of IoT, we aim to develop a system that ensures operational reliability and resilience of personalized and application-specific healthcare services to patients through a coordinated network of IoT applications and devices. In the following section, we discussed various components and workflow of our proposed solution.

\section{Proposed Solution}\label{sec:psolution}
To ensure the operational reliability of coordinated healthcare IoT devices and applications to deliver personalized healthcare services, we plan to develop a methodology that includes the following three components:
\begin{itemize}
    \item \textbf{Modeling:} To investigate various modeling techniques that help to develop an abstract but practical model of operational characteristics of IoT devices, applications, healthcare conditions, and standards. 
    \item \textbf{Reliability Analysis:} Based on the developed model, investigate various techniques to analyze the following  questions:
    \begin{itemize}
        \item How to establish reliable coordination among applications and devices?
        \item How to diagnose the right cause of the health condition?
        \item How to identify the right treatment that minimizes its side-effects on other health conditions?
    \end{itemize}
\item \textbf{Resilience Analysis:} To investigate techniques that ensure fail-safe operations of the devices and applications through continuous health monitoring.
\end{itemize}

%%%%%%%%%%%%%%%%%%%
\begin{table*}[ht]
\caption{Comparison of Various Smart healthcare Approaches }
\centering
 \begin{tabular}{|>{\centering\arraybackslash}m{3.3cm}|>{\centering\arraybackslash}m{1.5cm}|>{\centering\arraybackslash}m{1.5cm}|>{\centering\arraybackslash}m{4cm}|>{\centering\arraybackslash}m{4cm}|}
% \begin{tabular}{|c|c|c|c|c|}
    
\hline
\hline
\rowcolor{gray!20}

  {\bf{Healthcare Services}} & \textbf{Reliability} & \textbf{Resilience} & \textbf{Personalization (of individual health condition)}& \textbf{Personalization (of multiple health conditions)}\\
 \hline \hline
Smart healthcare~\cite{Minimed2016,lopez2015,yang2017,kouicem2018, khalil2020, letafatimehdi2020,sharma2016, Yanovsky2016} &\halfcirc[1.5ex] & \halfcirc[1.5ex] & \halfcirc[1.5ex] & \emptycirc[1.5ex]\\
\hline

 AI-based~\cite{HALEEM2021, zhan2020, Sloane:2020}  &\halfcirc[1.5ex] & \halfcirc[1.5ex] & \halfcirc[1.5ex]& \emptycirc[1.5ex]\\
\hline

 ML/DL-based~\cite{zeng2020,he2019}  &\halfcirc[1.5ex] & \halfcirc[1.5ex] & \halfcirc[1.5ex]& \emptycirc[1.5ex]\\
\hline

% & DL-based  &\ominus & \ominus & \ominus\\
% \hline

 CPHS (this work) & \fullcirc[1.5ex] & \fullcirc[1.5ex] & \fullcirc[1.5ex] & \fullcirc[1.5ex]\\
\hline
 
\end{tabular}
\label{table:comparisons}
\end{table*}
%%%%%%%%%%%%%%%%%%%

\subsection{Workflow of CPHS}\label{subsec:wcphs}
\begin{figure*}[!ht]
    \centering
    \includegraphics[scale=0.75]{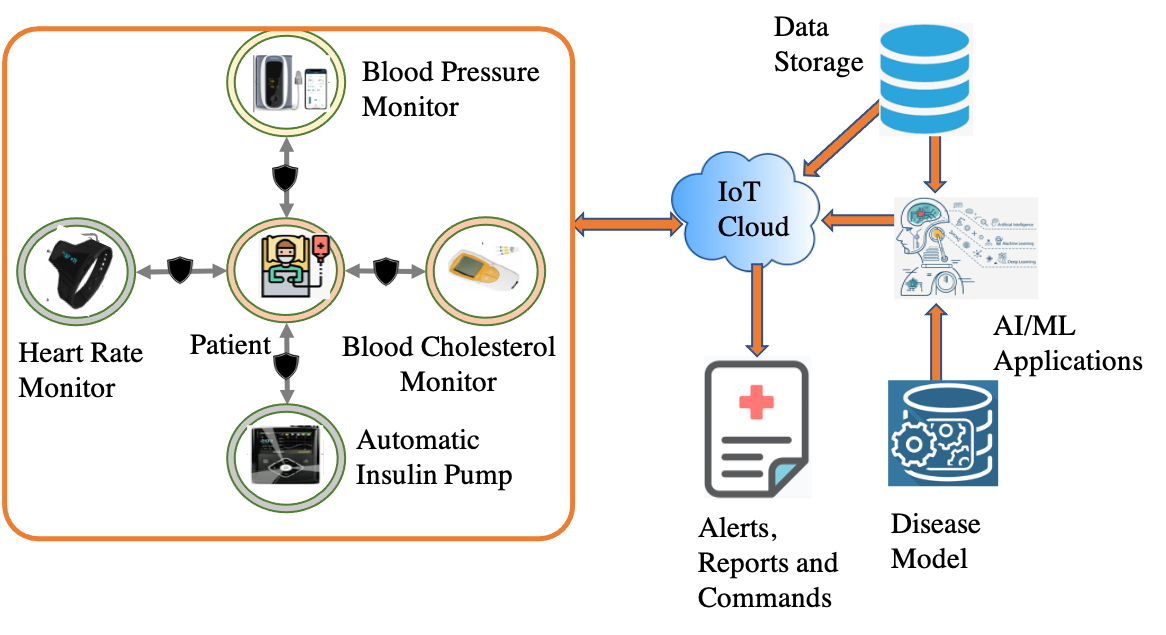}
    \caption{Workflow of Personalised HIoT}
    \label{fig:workflow}
\end{figure*}

The Workflow of the personalized HIoT System for comprehensive personalized healthcare is shown in Figure~\ref{fig:workflow}.
%%%%%%%%%%%%%%%%%%%%%%%%
%%%%%%%%%%%%%%%%%%%%%%%%%%%%%
Patient data from different monitoring devices like an insulin pump, blood pressure monitor, blood cholesterol monitor, and heart rate monitor is collected and transmitted at the IoT cloud where data is stored. Stored data is processed by using different AI techniques and disease models. However, several challenges are involved to develop the personalized HIoT System for comprehensive personalized healthcare.

To address these challenges, we aim to develop a solution that enables the modeling of health conditions based on their clinical/biological characteristics including clinical/biological dependencies among the conditions. Based on the models, we develop a personalized healthcare monitor that monitors various health conditions of a patient, and identifies any discrepancies among health conditions. Furthermore, the monitor provides clinical reasoning to rigorously identify the exact cause of the discrepancy. Once identified, the monitor can provide corresponding treatment to the discrepancy which can be later monitored to understand the effect of the treatment. Based on clinical reasoning, the monitor controls various health conditions of a patient remotely and automatically. Clinical reasoning makes the monitor self-aware, (i.e., it understands what it is monitoring), and thus supports healthcare services that are free of false diagnosis. Modeling health conditions is a challenging task as it requires identifying different clinical dependencies among various health conditions of a patient. However, these dependencies are typically non-trivial that require understanding of the underlying variable level biological processes associated with the conditions which are mostly a black box. To describe different levels of details about biological processes, we will introduce a modeling language that will allow modeling health conditions and their dependencies at an abstract but practical level of description. The model will be later used for mechanized reasoning about dependencies among the health conditions.
\subsection{Features of CPHS} \label{subsec:bcphs}
Our proposed methodology will not only realize personalized healthcare services based on their clinical characteristics but will also enable autonomous, automatic, and rigorous diagnosis of stealthy health conditions exploiting biological dependencies among various health conditions. Of course, we will address the key challenge to model clinical characteristics of health conditions that are highly declarative and abstract (e.g., health conditions with no specific clinical characteristics) on one hand, but are very low level (e.g., DNA and other details of a health condition) on the other hand. We aim to demonstrate the effectiveness of our methodology through its application to observe various health conditions (e.g., heart rate, diabetes, blood pressure, and cholesterol-related).

Consequently, we will develop a library of various health conditions (models) that can be later used in any healthcare computerized system to reason about healthcare services based on their clinical characteristics. The library will also help to learn other dependencies among various health conditions that are otherwise beyond the capabilities of medical practitioners. Moreover, based on the models, the newly identified dependencies will be explainable in a way that is understood by practitioners and machine.

Based on our results, we aim to integrate our method with following technologies:
\begin{itemize}
    \item Blockchain-based techniques to obtain integrity and reliability for personalized healthcare services.
    \item AI/ML techniques that will help to better understand health conditions by comparing data of different patients for similar health conditions and consequently proposing more personalized diagnosis and treatment, that will effectively and efficiently improve the health of patients.
\end{itemize}
As personalized healthcare and AI are evolving continuously, comprehensive biological knowledge, smart diagnostic techniques, and other factors like patient data from the medical images, and patient history will help to identify personalized therapies for individuals. AI is playing a vital role in different fields of healthcare. So far AI is used at Things, Communication, and Navigation layers for data collection, communication, and patient monitoring respectively. However, AI is not yet completely safe for diagnosis and treatment purposes due to its explainability and trust issues. Therefore, AI algorithms are not yet fully trusted in critical healthcare domains.

\section{Conclusion}\label{sec:conclusion}
We have introduced Healthcare 5.0, personalized healthcare services, and their key requirements that are classified into reliability, resilience, and personalized healthcare. Importantly, we have defined personalized healthcare services as a relationship of various health conditions of a patient based on the characteristics of their underlying biological process as well as their associated mechanical monitoring system. Based on the reference architecture of a modern IoT-based healthcare system, we have investigated the current efforts to address the defined key requirements. Finally, we have identified research gaps that need to be addressed to develop reliable and resilient personalized healthcare systems for the future. 

%\newpage
\bibliographystyle{IEEEtran}
% argument is your BibTeX string definitions and bibliography database(s)
\bibliography{paper}
\vspace{-1cm}
 \begin{IEEEbiography}[{\includegraphics[width=1in,height=1.25in,clip,keepaspectratio]{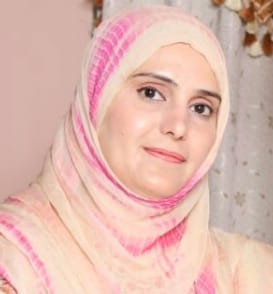}}]{Najma Taimoor}
 is a doctoral student at the Vienna University of Technology, Austria. Prior to that, she has completed her Masters in Healthcare and IT from the University of Naiveed Sciences, Austria with Distinction. Her Master's thesis was co-developed with the University of Surrey, the UK in the frame of the Erasmus grant. Her research interests include AI and IoT-based reliable healthcare systems.
 \end{IEEEbiography}
\vspace{-1cm}
 \begin{IEEEbiography}[{\includegraphics[width=1in,height=1.25in,clip,keepaspectratio]{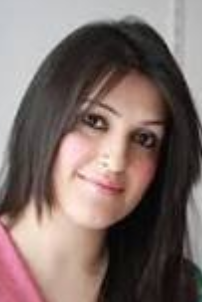}}]{Semeen Rehman} is currently with the Technische Universität Wien (TU Wien), Faculty of Electrical Engineering as a tenure-track Assistant Professor. In October 2020, she received her habilitation in the area of Embedded Systems from the Technische Universit\"at Wien (TU Wien). Before that, she was a Postdoctoral Researcher with the Technische Universität Dresden (TU Dresden) and Karlsruhe Institute of Technology (KIT), Germany, since 2015. In July 2015, she received her Ph.D. from Karlsruhe Institute of Technology (KIT), Germany. She has co-authored one book, multiple book chapters, and more than 50 publications in premier journals and conferences. Her main research interests include dependable systems, cross-layer design for error resiliency with a focus on run-time adaptations, emerging computing paradigms, such as approximate computing, hardware security, energy-efficient computing, embedded systems, MPSoCs, Internet of Things, and Cyber-Physical Systems. She has received the CODES+ISSS 2011 and 2015 Best Paper Awards, DATE 2017 Best Paper Award Nomination, several HiPEAC Paper Awards, Richard Newton Young Student Fellow Award at DAC 2015, and Research Student Award at KIT, in 2012. She has served as the TPC track chair for the ISVLSI 2020 and 2021 conference, and served as the TPC of multiple premier conferences on design automation and embedded systems (such as DAC, DATE, CASES, ASPDAC, VLSID). She has (co-)chaired multiple sessions at the DATE conference between 2019 and 2017. \end{IEEEbiography}

\EOD

\end{document}